\documentclass[10pt, conference, letterpaper]{IEEEtran}

\usepackage[pass]{geometry}

\usepackage{etex}
\usepackage{subcaption}
\usepackage[font=footnotesize]{caption}
\usepackage{algorithm}
\usepackage{amsfonts}
\usepackage{amsthm}
\usepackage{amsmath}
\usepackage{array}
\usepackage{bbm}
\usepackage[numbers,sort&compress]{natbib} 
\usepackage{dsfont}
\usepackage{epsfig}
\usepackage{float}
\usepackage[T1]{fontenc}
\usepackage{graphicx}
\usepackage{indentfirst}
\usepackage{multicol}

\usepackage{algorithmicx}
\usepackage{algpseudocode}

\usepackage{array}
\newcolumntype{L}[1]{>{\raggedright\let\newline\\\arraybackslash\hspace{0pt}}m{#1}}
\newcolumntype{C}[1]{>{\centering\let\newline\\\arraybackslash\hspace{0pt}}m{#1}}
\newcolumntype{R}[1]{>{\raggedleft\let\newline\\\arraybackslash\hspace{0pt}}m{#1}}

\usepackage{url}
\usepackage{multirow}
\usepackage{color}
\usepackage{blkarray} 
\usepackage{tikz}
\usetikzlibrary{calc,arrows,snakes,backgrounds}

\usepackage[export]{adjustbox} 
\usepackage{setspace}

\let\oldFootnote\footnote
\newcommand\nextToken\relax

\renewcommand\footnote[1]{%
    \oldFootnote{#1}\futurelet\nextToken\isFootnote}

\newcommand\isFootnote{%
    \ifx\footnote\nextToken\textsuperscript{,}\fi}

\usepackage{todonotes}


\usepackage[normalem]{ulem} 

\usepackage{float}

\usepackage[nopostdot,acronym,shortcuts,nonumberlist]{glossaries}
\newglossarystyle{longacro}{%
\setglossarystyle{long}
  {\end{longtable*}}%
}

\usepackage{booktabs}

\usepackage{tikz}
\usepackage{xcolor}
\definecolor{myblue}{HTML}{396AB1} 
\definecolor{myorange}{HTML}{DA7C30} 
\definecolor{mygreen}{HTML}{3E9651}
\definecolor{myred}{HTML}{CC2529}
\definecolor{mygray}{HTML}{535154}
\definecolor{mypurple}{HTML}{6B4C9A}
\definecolor{mymaroon}{HTML}{922428}
\definecolor{mymudgreen}{HTML}{948B3D}

\definecolor{mylblue}{HTML}{7293CB} 
\definecolor{mylorange}{HTML}{E1974C} 
\definecolor{mylgreen}{HTML}{84BA5B}
\definecolor{mylred}{HTML}{D35E60}
\definecolor{mylgray}{HTML}{535154}
\definecolor{mylpurple}{HTML}{6B4C9A}
\definecolor{mylmaroon}{HTML}{922428}
\definecolor{mylmudgreen}{HTML}{948B3D}
\newfont{\mycrnotice}{ptmr8t at 7pt}
\newfont{\myconfname}{ptmri8t at 7pt}

\IEEEoverridecommandlockouts

\DeclareMathSizes{10pt}{8pt}{6pt}{5pt} 	

\begin{document}

\def\sharedaffiliation{%
\end{tabular}
\begin{tabular}{c}}

\title{An Analytical Framework for mmWave-Enabled V2X Caching}

\author{
   \IEEEauthorblockN{Saeede Fattahi-Bafghi\IEEEauthorrefmark{1}, Zolfa Zeinalpour-Yazdi\IEEEauthorrefmark{1},  Arash Asadi\IEEEauthorrefmark{2}}

\\
    \IEEEauthorblockA{\IEEEauthorrefmark{1}Yazd University, Yazd, Iran}

    \IEEEauthorblockA{\IEEEauthorrefmark{2}Secure Mobile Networking lab (SEEMOO), Technische Universit\"{a}t Darmstadt, Darmstadt, Germany}}

\maketitle

\begin{abstract}
Autonomous vehicles will rely heavily on vehicle-to-everything (V2X) communications to obtain the large amount of information required for navigation and road safety purposes. This can be achieved through: $( i )$ leveraging millimeter-wave (mmWave) frequencies to achieve multi-Gbps data rates, and $(ii)$ exploiting the temporal and spatial correlation of vehicular contents to offload a portion of the traffic from the infrastructure via caching. Characterizing such a system under mmWave directional beamforming, high vehicular mobility, channel fluctuations, and different caching strategies is a complex task. In this article, we propose the first stochastic geometry framework for caching in mmWave V2X networks, which is validated via rigorous Monte XCarlo simulation. In addition to common parameters considered in stochastic geometry models, our derivations account for caching as well as the speed and the trajectory of the vehicles. Furthermore, our evaluations provide interesting design insights: $( i )$ higher base station/vehicle densities does not necessarily improve caching performance; $( ii )$ although using a narrower beam leads to a higher SINR, it also reduces the connectivity probability; and $( iii )$ V2X caching can be an inexpensive way of compensating some of the unwanted mmWave channel characteristic.
\end{abstract}

\section{Introduction}
\label{s:intro}

Autonomous driving is among the disruptive technological advances of the 21$^{\text{st}}$ century. However, contrary to prior automotive advancements, which essentially improved the mechanics, autonomous driving is about sensing, data processing, and connectivity. The latter is the most challenging task as it goes beyond on-vehicle enhancement, and it requires upgrading communication infrastructures to support data rates up to 20 Gbps~\cite{asadi2018fml, series2015imt}. {\it This data rate exceeds the advertised data rate of 5G ($\leq\!$~600 Mbps) and even the expected throughput of Beyond-5G networks in millimeter-wave (mmWave) bands ( $\leq\!$ 10 Gbps).}

Fortunately, the required information for autonomous vehicles has high temporal and spatial correlation, e.g., nearby vehicles need similar 3D high-resolution maps. {\it This allows effective use of caching mechanisms for traffic offloading to vehicles}. Therefore, content caching among vehicles is an effective method for reducing the overall load on the infrastructure. From a technical point of view, mmWave V2X caching is feasible since the vehicles have sufficient storage and computing power. Furthermore, vehicle-to-everything (V2X) communication (C-V2X in 3GPP terminology) enabled vehicles to communicate  directly with the infrastructure (V2I) and other vehicles (V2V). Indeed, after the Content Delivery Networks (CDNs), {\it autonomous driving will be the most popular use-case for caching}~\cite{hu2018vehicle,grewe2018caching}. There is, however, a major difference. The caching entities in CDNs are static whereas in autonomous driving, these entities (i.e., vehicles) are highly mobile, and the communication medium is shared.

{\bf Motivation.} Given the {\it mission-critical} nature of autonomous driving, robustness and reliability are the primary design goals for caching in mmWave V2X networks. In particular, cellular operators should ensure sufficient capacity for the delivery of contents in different network/traffic conditions. Unfortunately, {\it identifying the performance of these systems is complex due to the multitude of the variables in play; this includes the mobility pattern of the vehicles, the density of the vehicles and base stations (BSs), fluctuation of the communication channel, availability of caches, and the cache size}. This renders simulation-based approaches time consuming and unscalable for $ ( i ) $ network planning, $( ii ) $ evaluation of caching strategies, and $ ( iii )$ characterization of the network behavior under different system variables. Hence, there has been growing interest to use stochastic geometry which has been proven a powerful tool for large-scale performance analysis of complex networks, which should be otherwise studied through time-consuming simulations and expensive prototyping. This approach has gained significant popularity in recent years due to its tractability~\cite{singh2015tractable}. Using stochastic geometry for modeling rate and coverage probabilities in sub-6GHz cellular networks was initiated with the seminal work of Andrews {\it et al.} in~\cite{andrews2011tractable}. To date, this approach has been extended to a plethora of scenario including mmWave cellular networks to model coverage and capacity~\cite{bai2014coverage}, blockages~\cite{bai2013coverage}, initial access~\cite{alkhateeb2017initial}, as well as new application areas such as vehicular communication~\cite{giordani2018coverage}, aerial communications~\cite{banagar20193gpp}, and caching~\cite{MiticiGGB13}.


{\bf Challenge.} Stochastic geometry-based modeling of {\it caching in V2X mmWave networks in which directionality of antennas, blockages, and mobility are intertwined with content caching is a formidable task, which has not been addressed before}. This makes deriving the closed-form expressions for such a complex system without strong assumptions even more challenging. For example, many prior works simplify the derivations by neglecting the mobility and the moving trajectory of the vehicles, i.e., assuming vehicles' displacement is negligible in one slot.

\subsection{Methodology and contributions.}
From a methodological point of view, we follow the common stochastic geometry approach~\cite{singh2015tractable, andrews2011tractable, bai2013coverage, alkhateeb2017initial, bai2014coverage, giordani2018coverage, MiticiGGB13, KeshavarzianZT19, BlaszczyszynG15, LiuY17} by calculating the SINR expression as well as the probability of content retrieval. This expression is then used as the basis for computing the probability of coverage and further performance metrics. Then we depart from the state-of-the-art in Section~\ref{ss:con_analysis}, where we model the mobility/trajectory of the vehicles as well as associating contents to the requesting users. To the best of our knowledge, this is the first article considering caching in mmWave V2X networks. Consequently, we provide the {\it first analytical framework for cached-enabled mmWave V2X systems}. In addition to providing new design insights in this work, our stochastic geometry model can be used to analyze future caching policies. The main distinctions of this work are summarized as follows:


\begin{itemize}

\item We derive the coverage probability, whose derivation is similar to prior works with the exception that {\it we consider a more realistic association policy}. We allow a vehicle to request the content from nearby vehicles as long as they provide higher SINR than the closest BS. However, prior works on D2D caching simplified the association policy by allowing a single attempt to request contents from nearby users. The second attempt consists in retrieving the content from a BS.


\item We derive the probability that a vehicle leaves the beam coverage area within a time slot (i.e., beam sojourn probability). This is done by computing the beam width and the mobility speed/trajectory of vehicles. The majority of prior works assume either the vehicles are stationary or remain within the beam coverage during a time slot. {\it We consider both mobility and moving trajectory of the vehicles}. We compute the connectivity probability based on the expressions derived for the coverage probability and beam sojourn probability.

\item Our evaluations reveal interesting design insights: $( i )$~The SINR gain due to use of very narrow beams ($\sim\!\!10^\circ$) is dominated by the negative impact of the reduced beam coverage area as well as increased interference range, see Section~\ref{ss:speed}; $ ( ii )$~There exists an optimal value for the BS density, after which the connectivity probability is negatively affected by density, see Section~\ref{ss:density}. $ ( iii )~$ Unlike other caching scenarios, where higher density of cache-enabled users improves the performance, higher vehicular densities for mmWave V2X caching impact the system negatively, see Section~\ref{ss:cache}; and $( iv ) $~Increasing cache size appears to be an inexpensive and effective method for compensating the negative impact of interference and high vehicular velocity on the system, see Section~\ref{ss:cache}.

\end{itemize}

\section{Related Work}

Although stochastic geometric treatment of caching in cellular networks already exists, to the best of our knowledge, this is the first study for caching in mmWave V2X networks. For completeness, we report the state-of-the-art on caching and V2X communication separately.

{\bf Stochastic geometry for caching.} The authors in~\cite{MiticiGGB13} characterize the cost associated with data retrieval as opposed to the deployment of caches in wireless networks. Stochastic geometry has also been used to study different aspects of caching such as user satisfaction~\cite{BlaszczyszynG15}, spectral efficiency~\cite{LiuY17}, energy efficiency~\cite{KeshavarzianZT19}, and successful transmission probability in fog networks~\cite{WangLWL19}. The works in~\cite{MiticiGGB13,BlaszczyszynG15,LiuY17,KeshavarzianZT19,WangLWL19} focus mostly on edge caching. Cooperation within a heterogenous network among BSs, relays, and users for content dissemination, when a portion of users have caching ability has been studied in~\cite{yang2015analysis}. After the emergence of D2D communication, on-device caching has been widely studied. In~\cite{GolrezaeiDM14,GiatsoglouNKAV17}, the authors study the impact of D2D caching on enhancing network capacity. In particular, \cite{GolrezaeiDM14} studies how the capacity of the networks scales as the number of (interfering) D2D users increase, whereas~\cite{GiatsoglouNKAV17} focuses on analyzing offloading gain and content retrieval delay for their proposed D2D-Aware caching policy. In~\cite{AfshangDC16}, the optimal number of D2D transmitters is characterized using stochastic geometry to maximize the spectral efficiency. In~\cite{MaDCLMLV18}, the social relationship among users is used to derive an analytical expression on downlink capacity. In~\cite{ChenPK17}, stochastic geometry tools are used to derive closed-form approximate for cache-enabled D2D networks. Specifically, the authors investigate the cache placement under two objectives, namely, hit probability maximization and throughput maximization.
{\it None of the above consider mobility in their analysis.}

{\bf Stochastic geometry for V2X.} The body of work on stochastic modeling of V2X communication is very limited. In~\cite{ChoHC18}, the downlink outage probability for a typical vehicle is derived in an urban V2X scenario, where vehicles have group mobility and are clustered.  A stochastic geometry-based model for mmWave V2I communications is proposed in~\cite{tassi2017modeling}, in which the authors investigated the probability of blockage and expected throughput under varying vehicular densities in a multi-lane highway. The authors of~\cite{GiordaniRZZ18} derive analytical expressions for the probability of beam alignments and connection stability; this is the only work in which mobility is considered but only in a fixed trajectory and for V2I links. Uplink cellular V2X communication is analyzed in~\cite{sial2018stochastic}, where the probability of success is derived for both V2I and V2V links. However, mobility and moving trajectory are not considered.

In fact, the majority of the state-of-the-art do not account for the vehicular mobility or consider the mobility of the vehicle is limited within a one time frame, effectively assuming the impact of mobility is negligible. Being a key characteristic of V2X communication, we account for mobility in our analysis. To increase accuracy, we consider relevant parameters such as beam alignment, beamforming, and content availability, etc.

\section{System Model}
\label{s:sysmodel}
In this section, we describe the overall network architecture as well as channel model, beamforming model, and content delivery protocol. Specifically, we adapted the commonly used channel model in~\cite{singh2015tractable} and the sectorized antenna model in~\cite{Wang:ij}. We consider a cellular network consisting of a number of BSs and vehicular user equipments (V-UEs). BSs and V-UEs are spatially distributed according to two independent homogeneous Poison Point Processes (PPPs) $\phi^{b} $ and $\phi^{u} $ with densities $\lambda^{b} $ and $\lambda^{u} $, respectively, where it is assumed that $\lambda^{u} \gg \lambda^{b} $. All BSs and V-UEs are equipped with mmWave radio interfaces operating in $K_a$ band (26.5-40 GHz). We denote the number of required contents (e.g., 3D maps, real-time traffic information) by \textit{N} whose size is $S$ bits. The BSs have all \textit{N} contents, whereas V-UEs only have access to a limited set of contents, which are randomly distributed among V-UEs. 
%

\subsection{Channel Model}
\label{ss:channel_model}
Given the susceptibility of mmWave communication to blockage, it is imperative to distinguish between the probability of having a line-of-sight (LOS) or a non-line-of-sight (NLOS) links between the BSs and V-UEs. This probability depends on the structural characteristics of the environment (e.g., density of building). Intuitively, the probability of encountering a blockage for a 100 m link in a dense urban environment is much higher than a rural area. We denote ${\rm P}_{L^\nu}(r)$ as the probability that a mmWave link with length \textit{r} is LOS and $\nu$ $\in $\{b,u\} specifies whether the transmitter is a BS or a V-UE. According to the generalized blockage ball model~\cite{singh2015tractable}, we have: ${\rm P} _{L^\nu}(r)=e^{-a_{los}^{\nu}r} $,
where $a_{los}^{\nu}$ is determined by the average size and density of blockages for $\nu$ $\in $\{b,u\}. The probability of NLOS is equal to: ${\rm P}_{N^\nu}(r)=1-{\rm P}_{L^\nu}(r).$
For theoretical tractability, we assume that the network consists of four tiers $k=\{L^b,N^b,L^u,N^u\}$ which is indicative of LOS BS, NLOS BS, LOS V-UE and NLOS V-UE, respectively. We also consider the path loss model is as follows:

\begin{align}\label{Lk}
\setlength{\abovedisplayskip}{3pt}
\setlength{\belowdisplayskip}{3pt}
\ell _{k}(r) =r^{-\alpha _{k} }.e^{-\zeta_{k}r}\begin{array}{cc}{}\end{array}k=\{L^b,N^b,L^u,N^u\},
\end{align}
where $\alpha _{k}$ is the path loss exponent and $e^{-\zeta_{k}r}$ indicates atmospheric attenuation with attenuation constant $\zeta_{k}$ and \textit{r} is the distance between the requesting user and its closest serving node. Fading is assumed to follow Rayleigh distribution with an average of 1, i.e., $h\sim \exp(1)$.

\subsection{Directional Beamforming}
Operating in mmWave band, BSs and V-UEs are equipped with phased antenna arrays. The antenna arrays are assumed to perform directional beamforming where the main lobe is directed towards the dominant propagation path while smaller sidelobes direct energy in other directions. The main lobe gain and sidelobe gains of BSs are denoted by $g_{M}^b $ and $g_{m}^b$, respectively; and for V-UEs, the main lobe and sidelobe gains are denoted by $g_{M}^u$ and $g_{m}^u$, respectively. The beam direction of the interfering links is modeled as a uniform random variable on [0, 2$\pi$]. Therefore, the effective antenna gain $G_{i}^{\nu}$ from an arbitrary transmitter at the typical receiver is a discrete random variable described by:

\begin{equation}\label{Gi}
\setlength{\abovedisplayskip}{3pt}
\setlength{\belowdisplayskip}{3pt}
G_{i}^{\nu} =\left\{\begin{array}{c}\hspace{-1.5em} {g_{M}^\nu \begin{array}{cc} g_{M}^u & {\begin{array}{cc} {w.p.} & {p_{g_{M}^\nu g_{M}^u } =} \end{array}} \end{array}\frac{\psi^{\nu} }{2\pi } \frac{\psi^{u} }{2\pi } } \\ \hspace{-.6em}{g_{M}^\nu \begin{array}{cc} g_{m}^{u}  & {\begin{array}{cc} {w.p.} & {p_{g_{M}^\nu g_{m}^u } =} \end{array}} \end{array}\frac{\psi^{\nu} }{2\pi } \frac{1-\psi^{u} }{2\pi } } \\
\hspace{-.6em}{g_{m}^\nu \begin{array}{cc} g_{M}^u & {\begin{array}{cc} {w.p.} & {p_{g_{m}^\nu g_{M}^u } =} \end{array}} \end{array}\frac{1-\psi^{\nu} }{2\pi } \frac{\psi^{u} }{2\pi } } \\ {g_{m}^\nu \begin{array}{cc} {g_{m}^u } & {\begin{array}{cc} {w.p.} & {p_{g_{m}^\nu g_{m}^u } =} \end{array}} \end{array}\frac{1-\psi^{\nu} }{2\pi } \frac{1-\psi^{u} }{2\pi } } \end{array}\right.,
\end{equation}
where $\nu$ $\in $\{b,u\} specifies if the transmitter is a BS or a V-UE. $\psi^{\nu}$ is the beamwidth of the main lobe, and $p_{G_{i}^{\nu}}$ is the probability of having an antenna gain of $G_{i}^{\nu}$.
Note that the beamforming gain on the desired link is always $G_{0}^{\nu} = g_{M}^{\nu}g_{M}^u$; while $G_{i}^{\nu}$ in \eqref{Gi} provides the possible antenna array gains from interferers.

\subsection{Content Delivery Protocol}
\label{ss:content}
The content can be obtained from three different sources. A V-UE can potentially obtain the requested content from a nearby V-UE or directly from a BS. Alternatively, the V-UE may have already cached the content, i.e., having it locally. For analytical tractability, we define the following cases:

\begin{itemize}
\item \textbf{V2V}: The typical V-UE does not have the content in its cache, and the $n^{th}$ closest node with the requested content is a V-UE. Note that none of the $n-1$ closer V-UEs have cached the requested content and there is no BS closer than the $n^{th}$ V-UE to the typical V-UE.
\item \textbf{V2I}: The typical V-UE does not have the content in its cache, and the $n^{th}$ closest node is a BS, while $n\!-\!1$ closest nodes are V-UEs that do not have the content, thus the content is received from the nearest BS.
\item \textbf{Local case}: The typical V-UE has requested content in its cache, i.e., hit cache event.
\end{itemize}
The cache hit probability $p_{h}$ is given by $ p_{h}=\frac{K_{n}}{N} $,
where $K_{n}$ and $N$ are the cache size of V-UEs and total number of contents.
Note that following~\cite{andrews2011tractable, bai2014coverage, bai2013coverage, alkhateeb2017initial, di2015stochastic, giordani2018coverage}, the nearest node is the one which results in highest received power. The received power at the requesting V-UE from the $\textit{j}^{th}$ node in the $\textit{k}^{th}$ tier is defined as:

\begin{equation}
\setlength{\abovedisplayskip}{3pt}
\setlength{\belowdisplayskip}{3pt}
\label{Prkj}
P_{r_{k,j}} ={P_{t}}_{k} {g_{M}}_{k}\ell _{k}\left(\left\| r_{k,j} \right\| \right),\hspace{+1em}\begin{array}{cc}{}\end{array}k=\{L^b,N^b,L^u,N^u\},
\end{equation}
where ${P_{t}}_{k} $ and ${g_{M}}_{k} $ are the transmitted power and antenna's gain for $\textit{k}^{th}$ tier, and $r_{k,j} $ is the typical V-UE's distance from the $\textit{j}^{th}$ node in the $\textit{k}^{th}$ tier. The indices for the tier and the associated node for the typical user are:
\[\left\{k^{*},j^{*}\right\}=\begin{array}{ccc}\hspace{-.5em}{\arg }&\hspace{-.5em}{\max }&\hspace{-.5em}{P_{r_{kj}}}\end{array}\begin{array}{ccc} {} & {}\end{array}\hspace{-.5em}k=\{L^b,N^b,L^u,N^u\}.\]

\section{Coverage Analysis}
\vspace{-1mm}
\label{s:algo}

In this section, we first present the association rule between V-UE and its associated BS/V-UE and derive its corresponding probability. Then we compute the closed-form expressions for SINR coverage and rate coverage probabilities for a typical V-UE in a mmWave V2X network. These parameters will be the basis of the connectivity probability derived in the next section.

\subsection{Access Analysis}

In this section, we are going to calculate the probability that a typical V-UE will be associated to the $\textit{k}^{th}$ tier. If the typical V-UE is associated to the $\textit{k}^{th}$ tier, the received downlink SINR can be formulated as:
\begin{equation}\label{SINRk}
SINR_{k} =\frac{{P_{t}}_{k} G_{0}^{\nu} h_{k,0} \ell _{k} \left(\left\| r_{k^{*}j^{*}} \right\| \right)}{\sigma^{2}+I_{k}}\begin{array}{cc}{}\end{array} \!\!\!\! k=\{L^b,N^b,L^u,N^u\},
\end{equation}
where $G_{0}^{\nu}$=$g_{M}^{\nu}g_{M}^u$ is the effective antenna gain of the link between the typical V-UE and its associated BS/V-UE, $h$ is the small-scale fading coefficient from the associated BS/V-UE, $\sigma^{2}$ is the variance of the additive white Gaussian noise component and $I_{k}$ is the interference computed as:
\begin{equation}\label{Ik}
I_{k} \!\!=\!\! {\sum _{k} \sum _{j\in \phi _{k} \backslash j^{*} }\!\!\!\!\!\!{P_{t}}_{k} G_{i}^{\nu} h_{k,j} \ell _{k} \left(\left\| r_{k,j}\right\| \right)}\begin{array}{cc}{}\end{array} \!\!\!\! k\!=\!\{L^b,N^b,L^u,N^u\}.
\end{equation}
The effective antenna gains ($G_{i}^{\nu} $) was defined in~\eqref{Gi}.
According to PPP distribution, the probability of finding \textit{n} BSs/V-UEs in $\textit{k}^{th}$ tier in the area A with a radius of \textit{r} is given by:
\begin{equation} \label{fR1n}
\setlength{\abovedisplayskip}{3pt}
\setlength{\belowdisplayskip}{3pt}
{\mathbb P}(\begin{array}{ccc} \hspace{-.5em}{n} & \hspace{-.7em}{in}& \hspace{-.7em}{\phi_{k}} \hspace{-.7em}\end{array}\left|A=\pi r^{2} \right. )=\exp (-2\pi \lambda^{\nu} \Omega _{k} \left(r\right))\frac{(2\pi \lambda^{\nu}\Omega _{k} \left(r\right))^{n} }{n!},\nonumber
\end{equation}
where $\Omega_{k}(r)=\int _{0}^{r}{\rm P}_{k}(x )xdx$, and ${\rm P}_{k}(x)$ is given in Section~\ref{ss:channel_model} for $k$ $\in $ $\{L^b,N^b,L^u,N^u\}$. So the probability that the distance between the typical V-UE and closest BS/V-UE in $\textit{k}^{th}$ tier is larger than \textit{r} is:
\begin{equation} \label{FR12}
{\mathbb P}(R_{k} \ge r)={\mathbb P}(\begin{array}{ccc}{0} & \hspace{-.5em}{in}& \hspace{-.5em}{\phi_{k}} \hspace{-.5em} \end{array}\left|r)\right. =\exp (-2\pi \lambda^{\nu} \Omega _{k} \left(r\right)),
\end{equation}
where $R_{k}$ is the typical V-UE's distance from the closest BS/V-UE. Thus the probability that the distance between the typical V-UE and $\textit{n}^{th}$ closest BS/V-UE in $\textit{k}^{th}$ tier is larger than \textit{r} is equal to having at least $\textit{n}-1$ BSs/V-UEs in $\textit{k}^{th}$ tier in the area A with a radius of \textit{r} is:
\begin{align}\label{FRn2}
P\left(R_{k}^{\left(n\right)}>r\right)=& P\left(N<n-1\left|A=\pi r^{2} \right. \right)\nonumber \\
=&\sum _{i=0}^{n-1}\exp (-2\pi \lambda^{\nu} \Omega _{k} \left(r\right))\frac{(2\pi \lambda^{\nu} \Omega _{k} \left(r\right) )^{i}}{i!},
\end{align}
where $R_{k}^{(n)}$ is the typical V-UE's distance from the $\textit{n}^{th}$ closest BS/V-UE in the $\textit{k}^{th}$ tier. Therefore the CDF function for $R_{k}^{(n)}$ is as follows:
\begin{align}\label{FRn}
F_{R_{k}}^{\left(n\right)} \left(r\right)=&1-P\left(R_{k}^{\left(n\right)} >r\right) \nonumber \\
=&1-\sum _{i=0}^{n-1}\exp \left(-2\pi \lambda^{\nu} \Omega_{k} \left(r\right)\right)\frac{\left(2\pi \lambda^{\nu} \Omega _{k} \left(r\right)\right)^{i} }{i!},
\end{align}
The probability density function (PDF) of $R_{k} $ is given by:
\begin{equation} \label{fRn}
\!\! f_{R_{k}}^{\left(n\right)}\hspace{-.2em}\left(r\right)\hspace{-.2em}=\hspace{-.2em}2\pi r\lambda^{\nu} {\rm P}_{k} \left(r\right)\exp \left(-2\pi \lambda ^{\nu} \Omega_{k} \left(r\right)\right)\frac{\left(2\pi \lambda^{\nu} \Omega_{k}\left(r\right)\right)^{n-1} }{n-1!}.\!\!\!\!
\end{equation}
\textbf{Lemma 1}.
The association probability of a typical V-UE to $k^{th}$ tier in $n^{th}$ step of association, while in the previous  $n-1$ steps had access to LOS V-UEs $m$ times and NLOS V-UEs $n-m-1$ times and was not successful because of non-availability of the file of interest, is given by:
\begin{align*}
A_{k}^{\left(n,m\right)}\hspace{-.4em}=\hspace{-.3em}\int_{0}^{\infty}\hspace{-.8em}f_{R_{k}}^{\left(n_{k}\right)}\hspace{-.2em}\left(r\right)\hspace{-.8em}\prod _{i\in K\backslash k}\hspace{-.5em}\left[F_{R_{i} }^{\left(n_{i}\right)}\hspace{-.2em}\left(\Lambda _{k,i}\left(r\right)\right)\hspace{-.2em}-\hspace{-.2em}F_{R_{i}}^{\left(n_{i}-1\right)} \hspace{-.2em}\left(\Lambda _{k,i}\left(r\right)\right)\right]\hspace{-.2em}dr,
\end{align*}
where $n_{L^b}=n_{N^b}=1$, $n_{L^u}=m+1$ and $n_{N^u}=n-m$. $\Lambda_{k,i}(r)=\ell^{-1}_{i}((\frac{{P_{t}}_{k}g_{M_{k}}}{{P_{t}}_{i}g_{M_{i}}})\ell_{k}(r))$ where $\ell_{k}(r)=r^{-\alpha_{k}}e^{-\zeta_{k}r}$ is path loss model and ${\ell^{-1}}_{k}(r)$ is Lambert-W function. Note $F_{R_{i}}^{\left(0\right)}\left(r\right)=0$.
\begin{proof}
See Apendix A.
\end{proof}
\textbf{Corollary 1}. Denote $X_{k}$ as the distance between a typical V-UE to $k^{th}$ tier in $n^{th}$ step of association, where in the previous $n-1$ steps had access to  LOS V-UEs $m$ times and  NLOS V-UEs $n-m-1$ times. The PDF of $X_{k}$ is given by:
\begin{align*}
f_{X_{k}}^{\left(n,m\right)}(x)\hspace{-.2em}=\hspace{-.2em}\frac{f_{R_{k} }^{\left(n_{k}\right)} \hspace{-.2em}\left(r\right)}{A_{k}^{\left(n,m\right)} }\hspace{-.7em}\prod _{i\in K\backslash k}\hspace{-.5em}[F_{R_{i} }^{\left(n_{i}\right)} \hspace{-.2em}\left(\Lambda _{k,i}\left(r\right)\right)\hspace{-.2em}-\hspace{-.2em}F_{R_{i} }^{\left(n_{i}-1\right)}\hspace{-.2em} \left(\Lambda _{k,i}\left(r\right)\right)].
\end{align*}
\begin{proof}
See Apendix B.
\end{proof}
\textbf{Lemma 2}. Having the association probability, we can calculate the probability of the three content retrieval case (see Section~\ref{ss:content}) as follows
\begin{align} \label{Pcase}
    P_{i}& =
    \begin{cases}
       p_{h}, &  i=Local\\
       \sum _{k=\{L^b,N^b\}}\sum _{n=1}^{\infty}\sum _{m=0}^{n-1}P_{V2I,k}^{(n,m)}, &  i=V2I\\
       \sum _{k=\{L^u,N^u\}}\sum _{n=1}^{\infty}\sum _{m=0}^{n-1}P_{V2V,k}^{(n,m)}, &  i=V2V
    \end{cases}
\end{align}
where $p_{h}$ is cache hit probability and
\begin{align} \label{Pnm}
 \!\!\!\!\!   P_{i,k}^{(n,m)}&\! =\!
    \begin{cases}
       (1-p_{h})^{n}A_{k}^{(n,m)}, &k=\{L^b,N^b\},i=V2I\\
       (1-p_{h})^{n}p_{h}A_{k}^{(n,m)}. &k=\{L^u,N^u\},i=V2V
    \end{cases}
\end{align}

\subsection{SINR Coverage Probability}

 The SINR coverage probability $SC_{k} \left(\tau \right)$ is defined as the probability that the received SINR is larger than a certain threshold $\tau >0$, when the typical user is associated with a BS/V-UE from the $\textit{k}^{th}$ tier.\\
\textbf{Lemma 3}. The SINR coverage probability of the $\textit{k}^{th}$ tier is given by:
 \begin{align} \label{SCk}
SC_{k} \left(\tau \right) &=p(SINR_{k} >\tau |K=k) \nonumber \\
				   &=p\left(\frac{{P_{t}}_{k}G_{0}^{\nu} h \ell _{k} \left(\left\| x \right\| \right)}{\sigma^{2} +I}>\tau \left|K=k\right. \right)  \nonumber \\
				   &=\int _{0}^{\infty }\hspace{-.5em}\exp\hspace{-.2em}\left(\hspace{-.2em}\frac{-\tau \sigma^{2} }{T_{k}\hspace{-.2em}\left(x\right)}\hspace{-.2em} \right)\hspace{-.2em}L_{I}\hspace{-.2em}\left(\hspace{-.2em}\frac{\tau}{T_{k}\hspace{-.2em} \left(x\right)}\hspace{-.2em}\right)\hspace{-.2em}f_{X_{k}}\hspace{-.2em}\left(x\right)dx,
\end{align}
where $T_{k}(x)= {P_{t}}_{k}G_{0}^{\nu}\ell _{k} \left(\left\| x \right\| \right)$ and $k=\{L^b,N^b,L^u,N^u\}$. $L_{I}(.)$ is the Laplace transforms of interference $I$ given by
\begin{align} \label{LI}
L_{I}\left(s\right)&={\exp \left(\hspace{-.2em}-2\pi \lambda^{\nu}\sum _{i}p_{G_{i}^{\nu}}\sum _{k}W_{k}(\alpha_{k},\zeta_{k},s{P_{t}}_{k}G_{i}^{\nu})\right)},
\end{align}
where
\begin{align} \label{LI}
W_{k}(\alpha_{k},\zeta_{k},\eta_{k})&=\int _{0}^{\infty }\frac{r{\rm P}_{k}\left(r\right)}{1+\eta_{k}^{-1}r^{\alpha_{k}}e^{\zeta_{k}r}}dr.
\end{align}
Details of the proof are presented in Appendix C.\\
In the following SINR coverage probability is obtained for all three cases separately.

\paragraph{ SINR coverage of V2I case}
The typical V-UE takes content from the closest LOS or NLOS BS, hence:
\begin{align} \label{SCV2Ik}
\setlength{\abovedisplayskip}{3pt}
\setlength{\belowdisplayskip}{3pt}
SC_{V2I,k}^{(n,m)} \left(\tau \right)&\hspace{-.2em}=\hspace{-.2em}\int _{0}^{\infty }\hspace{-.5em}\exp \hspace{-.2em}\left(\hspace{-.2em}\frac{-\tau \sigma^{2} }{T_{k}\hspace{-.2em} \left(x \right)} \hspace{-.2em}\right)\hspace{-.2em}L_{I}\hspace{-.2em} \left(\hspace{-.2em}\frac{\tau }{T_{k} \hspace{-.2em}\left(x \right)}\hspace{-.2em}\right)\hspace{-.2em}f^{(n,m)}_{X_{k} }\hspace{-.2em} \left(x\right)dx,
\end{align}
where $k=\{L^b,N^b\}$. The total SINR coverage for V2I is:
\begin{equation} \label{SCV2I}
\setlength{\abovedisplayskip}{3pt}
\setlength{\belowdisplayskip}{3pt}
SC_{V2I} \left(\tau \right)=\hspace{-.5em}\sum _{k=\{L^b,N^b\}}\sum _{n=1}^{\infty}\sum _{m=0}^{n-1}\frac{P_{V2I,k}^{(n,m)}}{P_{V2I}}SC_{V2I,k}^{(n,m)} \left(\tau \right).
\end{equation}

\paragraph{SINR coverage of V2V case}
The typical V-UE receives the content from a LOS or a NLOS V-UE in $n^{th}$ step of the association. The SINR coverage probability of the $\textit{k}^{th}$ tier ($k=\{L^u,N^u\}$) in the $n^{th}$ step is given by:
\begin{align} \label{SCV2Vk}
\setlength{\abovedisplayskip}{3pt}
\setlength{\belowdisplayskip}{3pt}
SC_{V2V,k}^{(n,m)}\hspace{-.2em}\left(\tau \right)&\hspace{-.2em}=\hspace{-.2em}\int _{0}^{\infty }\hspace{-.5em}\exp\hspace{-.2em}\left(\hspace{-.2em}\frac{-\tau \sigma^{2}}{T_{k}\hspace{-.2em}\left(x\right)}\hspace{-.2em} \right)\hspace{-.2em}L_{I}\hspace{-.2em}\left(\hspace{-.2em}\frac{\tau }{T_{k}\hspace{-.2em} \left(x\right)}\hspace{-.2em}\right)\hspace{-.2em}f^{(n,m)}_{X_{k}}\hspace{-.2em}\left(x\right)dx,
\end{align}
so
\begin{equation} \label{SCV2V}
SC_{V2V} \left(\tau \right)=\hspace{-.5em}\sum _{k=\{L^u,N^u\}}\sum _{n=1}^{\infty}\sum _{m=0}^{n-1}\frac{P_{V2V,k}^{(n,m)}}{P_{V2V}}SC_{V2V,k}^{(n,m)} \left(\tau \right).
\end{equation}

\paragraph{ SINR coverage of Local case}
In this case, the content is locally available. Therefore, $SC_{Local}\left(\tau \right)=1.$

The average SINR coverage probability is obtained by the product of the probability of each case in its SINR coverage probability as:
\begin{equation} \label{SC}
\setlength{\abovedisplayskip}{3pt}
\setlength{\belowdisplayskip}{3pt}
SC\left(\tau \right)=\sum _{i}P_{i} SC_{i} \left(\tau \right),
\end{equation}
where $P_{i}$ for $i\in\{V2I, V2V, Local,\}$ are defined in \eqref{Pcase}.

\subsection{ Rate Coverage Probability}

In this subsection, we derive the rate coverage probability for the typical V-UE. Since rate characterizes the throughput observed in the network, it is an important performance metric like SINR. Similar to SINR coverage probability, the rate coverage probability $RC^{(n,m)}_{i,k}(\rho )$ is defined as the probability that the rate is larger than a certain threshold $\rho>0$ when the typical V-UE is associated to a BS/V-UE in the $\textit{k}^{th}$ tier and in $n^{th}$ step of association, where in the previous  $n-1$ steps had access to $m$ LOS V-UEs and $n-m-1$ NLOS V-UEs. Hence, the total rate coverage $RC_{i} \left(\rho \right)$ of the network for the $\textit{i}^{th}$ case is:
\begin{equation} \label{RCi}
RC_{i}\left(\tau \right)=\sum _{k}\sum _{n=1}^{\infty}\sum _{m=0}^{n-1}P_{i,k}^{(n,m)}RC_{i,k}^{(n,m)} \left(\tau \right).
\end{equation}
where the conditional rate coverage probability $RC^{(n,m)}_{i,k}$ is calculated in terms of SINR coverage probability as:
\begin{align} \label{RCnmik}
RC^{(n,m)}_{i,k} (\rho )&={\mathbb P}(Rate^{(n,m)}_{k} >\rho )\nonumber \\
&={\mathbb P}  (\frac{W_{k} \log _{2} (1+SINR^{(n,m)}_{k} )}{N^{(n,m)}_{k} } >\rho ) \nonumber \\
&={\mathbb P}  (SINR^{(n,m)}_{k} >2^{\frac{\rho N^{(n,m)}_{k} }{W_{k} } } -1)\nonumber \\
&=SC^{(n,m)}_{i,k}(2^{\frac{\rho N^{(n,m)}_{k} }{W_{k}}}-1),
\end{align}

where $SC^{(n,m)}_{i,k}\left(.\right)$  is the SINR coverage probability of the $\textit{k}^{th}$ tier and $N^{(n,m)}_{k}$ and $W_{k}$ are referred to the load and total available resources in the $\textit{k}^{th}$ tier. The load $N^{(n,m)}_{k}$ indicates the total number of V-UEs served by the serving node in the $\textit{k}^{th}$ tier in $n^{th}$ step of association, is as follows:

\begin{equation} \label{N}
\setlength{\abovedisplayskip}{3pt}
\setlength{\belowdisplayskip}{3pt}
N^{(n,m)}_{k} =1+\frac{P^{(n,m)}_{i,k}\lambda ^{u}}{\lambda _{k}},
\end{equation}
where $\lambda _{k}=\lambda^{\nu}\Omega_{k}(\infty)$. Considering the occurrence probability of each three cases defined in (11), the total rate coverage probability is:

\begin{equation} \label{RC}
\setlength{\abovedisplayskip}{3pt}
\setlength{\belowdisplayskip}{3pt}
RC\left(\rho \right)=\sum _{i}P_{i} RC_{i} \left(\rho \right).
\end{equation}

\begin{figure}[t!]
	\centering
	\begin{subfigure}[t]{0.4\textwidth}
                	\resizebox{\linewidth}{!} {\input{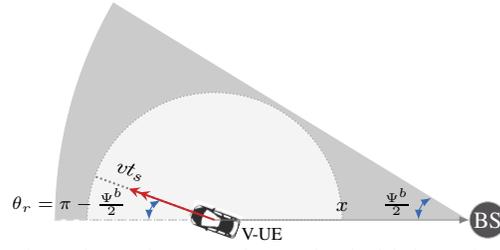}}
	\end{subfigure}
	\vspace{-3mm}	
	\caption{The motion angle V-UE to its associated BS is larger than $\pi -\frac{\psi^{b}}{2}$.}
	\label{fig:case_1}
	\vspace{-5mm}
\end{figure}

\section{Connectivity Analysis}
\vspace{-1mm}
 \label{ss:con_analysis}
To establish a communication link, the transmitter and the receiver should align their beams. This is typically performed in the so-called beam alignment procedure, which is executed periodically at the beginning of each time slot $t_{s}$. After beam alignment, the communication is initiated.
However, communication is only successful if the SINR is greater than a predefined threshold $\tau$. We refer to this as the SINR coverage which is computed before.
The typical V-UE could remain in the beam coverage of its associated BS within the whole time slot and maintain the connection; or it could lose the beam alignment and get disconnected. Consequently, the connectivity during the time slot requires the V-UE to be associated to a BS/V-UE at the beginning of the slot, have an SINR$>\tau$, and do not leave the beam coverage area (i.e., maintain the beam alignment).  We denote the latter by beam sojourn probability which is computed as:

\begin{figure*}[t!]
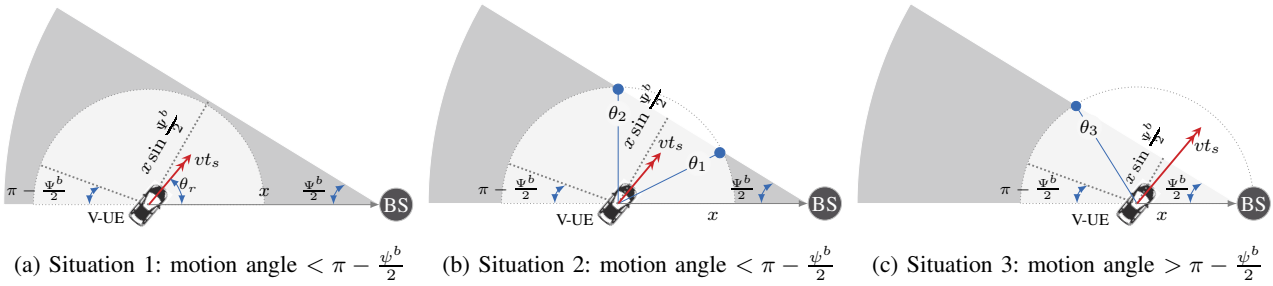

	\begin{subfigure}[t]{0.33\textwidth}
                	\resizebox{\linewidth}{!} {\input{./figs/mode2_d2b}}
                	\caption{Situation 1: motion angle $< \pi -\frac{\psi^{b}}{2}$}
                 \label{fig:case_2_1}
	\end{subfigure}	
	\begin{subfigure}[t]{0.33\textwidth}
                	\resizebox{\linewidth}{!} {\input{./figs/mode2_d2d}}
                	\caption{Situation 2: motion angle $< \pi -\frac{\psi^{b}}{2}$}
                 \label{fig:case_2_2}
	\end{subfigure}		
	\begin{subfigure}[t]{0.33\textwidth}
                	\resizebox{\linewidth}{!} {\input{./figs/mode3_d2b}}
                	\caption{Situation 3: motion angle $>\pi -\frac{\psi^{b}}{2}$}
                 \label{fig:case_2_3}
	\end{subfigure}			
	\caption{Different cases for the relative movements of V-UE with respect to its associated BS.}
	\label{fig:beam-sojourn}
\end{figure*}

\subsection{Beam Sojourn Probability}\label{Sec:Beam_Sojourn}
Here we derive the beam sojourn probability $B_{i,k}(x)$ which is the expected probability that a V-UE  of case $i$ ($i\in\{V2I, V2V,\}$),  associated to the $k^{th}$ tier, remains in the coverage of its associated BS/V-UE with a distance of $x$ from it. We first analyze the V2I communications and then  extend the analysis to the case of a V2V link, where the requesting V-UE receives the content from another V-UE.

For the V2I communications, where the trajectory of the V-UE has an angle $\theta$ with the bisector of associated BS's beam with the width of $\psi^{b}$, two scenarios may happen\footnote{The motion angle in our model is $[0-\pi]$. Nonetheless, we can easily adapt the model to $[\pi-2\pi]$. }:

{\bf Always in beam coverage.} If $\theta >\pi -\frac{\psi^{b}}{2} $, then for each distance $x$, speed $v$, and time slot $t_{s}$, the V-UE remains in the range of the associated BS, see Fig.~\ref{fig:case_1}. So $B_{V2I} =1$.

{\bf Conditionally in beam coverage.} If $\theta \le \pi-\frac{\psi^{b}}{2} $, then  $B_{V2I,k}\left(x\right)$ is computed under the following three situations, which are graphically depicted in Fig.(~\ref{fig:beam-sojourn}).

{\bf Situation 1.} If $x>\frac{vt_{s} }{\sin (\frac{\psi^{b}}{2} )}$, then V-UE does not leave the beam coverage area of the associated BS in time slot $t_{s}$, see Fig.~\ref{fig:case_2_1}. So again $B_{V2I,k}\left(x\right)=1$.

{\bf Situation 2.} If $vt_{s} \le x\le \frac{vt_{s} }{\sin (\frac{\psi^{b} }{2} )}$, then V-UE leaves the beam coverage area of the associated BS for $[\theta _{1} ,\theta _{2} ]$, where $\theta _{1} =\sin ^{-1}(\frac{x\sin(\frac{\psi^{b} }{2})}{vt_{s} })-\frac{\psi^{b} }{2}$ and $\theta _{2}=\pi -\sin ^{-1}(\frac{x\sin(\frac{\psi^{b} }{2})}{vt_{s} })-\frac{\psi^{b} }{2}$, see Fig.~\ref{fig:case_2_2}.  Hence $B_{V2I,k}\left(x\right)$ in this case is expressed as:

\begin{align}
\!\!\!\!\!\!\!\!\!\! B_{V2I,k}\left(x\right)\!&=P(\theta <\theta _{1} ,\theta >\theta _{2}|\theta<\pi-\frac{\psi^{b}}{2})\nonumber\\
&=\! \frac{\pi \!-\! \frac{\psi^{b} }{2} \!-\! \left(\theta _{2}  \!-\! \theta _{1} \right)}{\pi -\frac{\psi^{b} }{2}}\hspace{-.2em} =\hspace{-.2em}\frac{2\sin ^{-1}\hspace{-.2em}(\frac{x\sin(\frac{\psi^{b} }{2} )}{vt_{s} })-\frac{\psi^{b}}{2} }{\pi -\frac{\psi^{b}}{2} }.
\label{eq:V2I-2}
\end{align}
{\bf Situation 3.} If $x<vt_{s}$, then V-UE leaves the beam coverage area of the associated BS for $[0,\theta _{3}]$, where $\theta _{3} =\pi -\sin ^{-1}(\frac{x\sin(\frac{\psi^{b} }{2})}{vt_{s} })-\frac{\psi^{b} }{2} $. Hence $B_{V2I,k}\left(x\right)$ in this case is given by
\begin{align}
B_{V2I,k}\left(x\right) &=P(\theta >\theta _{3} ,|\theta <\pi -\frac{\psi^{b} }{2})\nonumber\\
&=\frac{\pi -\frac{\psi^{b} }{2} -\theta _{3} }{\pi -\frac{\psi^{b} }{2} }=\frac{\sin ^{-1} (\frac{x\sin \left(\frac{\psi^{b} }{2} \right)}{vt_{s} } )}{\pi -\frac{\psi^{b} }{2} }.
\label{eq:V2I-3}
\end{align}

The probability of three above mentioned  situations is expressed in the form of a piecewise function as follows:
\begin{align}
\!\!\!\!\!\! B_{V2I,k}\left(x\right)\!=\!
    \begin{cases}
       \frac{\sin ^{-1} (\frac{x\sin (\frac{\psi^{b} }{2} )}{vt_{s} } )}{\pi - \frac{\psi^{b} }{2}}, &x<vt_{s} ,\\
       \frac{2\sin ^{-1} (\frac{x\sin (\frac{\psi^{b}}{2})}{vt_{s} } )-\frac{\psi^{b}}{2} }{\pi -\frac{\psi^{b}}{2} },\hspace{0.8em} vt_{s}\le\!\!\!\!\!&x\le \frac{vt_{s} }{\sin (\frac{\psi^{b}}{2} )} \\
       1, &x>\frac{vt_{s} }{\sin (\frac{\psi^{b}}{2} )}
    \end{cases}
    \label{BV2Ik}
\end{align}

 We  follow the same approach as above to derive the sojourn probability for the V2V scenario. The main difference to V2I scenario is that instead of having a fixed speed $v$, we now deal with the relative speed of two V-UEs $v'=\left|2v\cos \left(\beta \right)\right|$ where $\beta =\frac{\theta _{r} -\theta _{t} }{2} $, and $\theta _{r}$ and $\theta _{t} $ are the receiver  and the transmitter angles, respectively. Also $\theta =\frac{\theta _{r}+\theta _{t} }{2} $ is the average motion angle. Given the uniform distribution of $\theta _{r}$ and $\theta _{t} $, the probability distribution functions of $\beta$ and  $\theta $ are as follows:
\begin{align}
    f_{\beta } \left(\beta \right)& =
    \begin{cases}
       \frac{\pi +\beta }{\pi ^{2}} ,&  -\pi <\beta <0,\\
       \frac{\pi -\beta }{\pi ^{2} },& 0<\beta <\pi
    \end{cases}
\end{align}
\begin{align}
    f_{\theta } \left(\theta\right)& =
    \begin{cases}
       \frac{\theta }{\pi ^{2} } ,& 0<\theta <\pi,\\
       \frac{2\pi -\theta }{\pi ^{2} } ,& \pi <\theta <2\pi
    \end{cases}
\end{align}
Substituting the new speed and new distributions of the angles in above equations, $B_{V2V,k}\left(x,\beta\right)$ is calculated as \\

\begin{align}
\label{BV2Vk}
B_{V2V,k}\left(x,\beta\right)&=\\
&\hspace{-5.1em}\left\{\begin{array}{c} {\hspace{-1.4em}\frac{\sin\hspace{-.2em}^{-1}\hspace{-.2em} \left(\hspace{-.2em}\frac{x\sin\left(\frac{\psi^{u} }{2} \right)}{2vt_{s} \cos (\beta )}\hspace{-.2em} \right)\hspace{-.2em}\left(\hspace{-.2em}2\pi -\psi^{u} -\sin\hspace{-.2em}^{-1}\hspace{-.2em}\left(\hspace{-.2em}\frac{x\sin \left(\frac{\psi^{u} }{2}\right)}{2vt_{s} \cos (\beta )} \hspace{-.2em}\right)\hspace{-.2em}\right)}{\left(\pi-\frac{\psi^{u} }{2} \right)^{2}}, \begin{array}{cc}{}{\hspace{.3em}x\hspace{-.2em}<\hspace{-.2em}2vt_{s}\hspace{-.1em}\cos(\beta )}\end{array}} \nonumber\\
{\hspace{-1.5em}\frac{(\hspace{-.1em}\frac{\psi^{{u}}}{2}\hspace{-.1em})^{2}\hspace{-.1em}+\hspace{-.1em}2\left(\hspace{-.1em}\pi\hspace{-.1em} -\hspace{-.1em}\psi^{u} \hspace{-.1em}\right)\hspace{-.1em}\sin \hspace{-.2em}^{-1} \hspace{-.2em}\left(\hspace{-.2em}\frac{x\sin \left(\frac{\psi^{u} }{2}\hspace{-.2em} \right)}{2vt_{s} \cos (\beta )} \hspace{-.2em}\right)}{\left(\pi -\frac{\psi^{u} }{2} \right)^{2}}, \begin{array}{cc} {} \hspace{+.5em}{2vt_{s}\hspace{-.1em} \cos (\beta )\hspace{-.2em}\le\hspace{-.2em} x\hspace{-.2em}\le\hspace{-.2em} \frac{2vt_{s} \hspace{-.1em}\cos (\beta )}{\sin (\frac{\psi^{u} }{2} )} } \end{array}} \nonumber\\
 {1.\begin{array}{ccc} {} & {} & {\begin{array}{cc} {\begin{array}{ccc} {\begin{array}{ccc} {} & {} & {} \end{array}} & {} & {} \end{array}} {\hspace{7.3em}x\hspace{-.2em}>\hspace{-.2em}\frac{2vt_{s} \cos (\beta )}{\sin (\frac{\psi^{u} }{2} )} } \end{array}} \end{array}} \end{array}\right.
\end{align}

\subsection{Connectivity Probability}
As stated before, a connection establishment  depends on two major factors, remaining in the beam area  and having a good link quality. Considering these two factors analyzed as beam sojourn probability and SINR coverage probability in the previous sections, we can derive the connectivity probability.
The connectivity probability of a V-UE of $\textit{k}^{th}$ tier, associated a BS/V-UE in the $n^{th}$ step (while in the previous  $n-1$ steps had access to  LOS V-UEs $m$ times and  NLOS V-UEs $n-m-1$ times and was not successful because of unavailability of the file of interest), is equal to:
\begin{align}\label{BnmV2Ik}
PC^{(n,m)}_{V2I,k}&=\int_{0}^{\infty }SC^{(n,m)}_{V2I,k}(\tau,x)B_{V2I,k}(x)dx\nonumber\\
PC^{(n,m)}_{V2I,k}&=\hspace{-.1em}\frac{\psi^{b} }{2\pi }SC^{(n,m)}_{V2I,k} \left(\tau \right)+\hspace{-.1em}\frac{2\pi -\psi^{b} }{2\pi }  \hspace{-.1em}\nonumber\\
&\hspace{-4em}\int _{0}^{\infty }\hspace{-.5em}\exp  \hspace{-.2em}\left( \hspace{-.1em}\frac{-\tau \sigma^{2} }{T_{k} \left(x\right)}  \hspace{-.1em}\right) \hspace{-.2em}L_{I}\hspace{-.2em}\left( \hspace{-.1em}\frac{\tau }{T_{k} \left(x\right)} \hspace{-.1em}\right) \hspace{-.1em}B_{V2I,k}(x)f^{(n,m)}_{X_{k}}(x)dx.
\end{align}
Thus the total connectivity probability for V2I scenario is:
\begin{equation} \label{PCV2I}
PC_{V2I} \left(\tau \right)=\sum _{k=\{L^b,N^b\}}\sum _{n=1}^{\infty}\sum _{m=0}^{n}\frac{P_{V2I,k}^{(n,m)}}{P_{V2I}}PC_{V2I,k}^{(n,m)} \left(\tau \right).
\end{equation}
Similarly the probability of connectivity in the $\textit{k}^{th}$ tier for  the association of step $n$ is given by:
\begin{align} \label{BnmV2Vk}
PC^{(n,m)}_{V2V,k}&=\int _{0}^{\infty }SC^{(n,m)}_{V2V,k} (\tau,x)B_{V2V,k}(x)dx\nonumber\\
&=\frac{\left(2\pi -\frac{\psi^{u} }{2} \right)\frac{\psi^{u} }{2} }{\pi ^{2} }SC^{(n,m)}_{V2V,k} \left(\tau \right)+\frac{\left(\pi -\frac{\psi^{u} }{2} \right)^{2} }{\pi ^{2} }\nonumber\\
&\hspace{-4.5em}\int_{\hspace{-.3em}-\pi}^{\pi }\hspace{-.4em}\int _{0}^{\infty }\hspace{-1.2em}\exp \hspace{-.2em}\left(\hspace{-.2em}\frac{-\tau \sigma^{2} }{T_{k}\hspace{-.2em}\left(x\right)} \hspace{-.2em}\right)\hspace{-.3em}L_{I}\hspace{-.3em}\left(\hspace{-.2em}\frac{\tau }{T_{k}\hspace{-.2em}\left(x\right)}\hspace{-.2em}\right)\hspace{-.3em}B_{V2V,k}\hspace{-.1em}(x,\beta)\hspace{-.1em}f^{(n,m)}_{X_{k}}\hspace{-.1em}(x)\hspace{-.1em}f_{\beta }\hspace{-.2em}\left(\beta\right)\hspace{-.1em}dxd\beta\nonumber,
\end{align}
The above probability $PC^{(n,m)}_{V2V,k}$ can be written as:
\begin{equation} \label{PCV2V}
PC_{V2V}\left(\tau \right)=\sum _{k=\{L^u,N^u\}}\sum _{n=1}^{\infty}\sum _{m=0}^{n-1}P_{V2I,k}^{(n,m)}PC_{V2V,k}^{(n,m)} \left(\tau \right).
\end{equation}
Based on \eqref{PCV2I} and \eqref{PCV2V}, the total connectivity probability for a typical V-UE is computed as follows:

\begin{equation}
\label{PCT}
PC=\sum _{i\in\{V2I, V2V, Local,\}}P_{i}PC_{i}.
\end{equation}
where $P_{i}$ is the probability of the occurrence of case $i$ and $PC_i(.)$ is its connectivity probability.

\section{Throughput and Delay Analysis}

In this section, we derive the total average throughput $T$ for the typical V-UE and also the experienced delay in receiving  the requested content of size $S$. To this aim, in Lemma 4 and Lemma 5 the average rate ($E[{t}^{(n,m)}_{i,k}]$) and  the average connection time $E[{t}^{(n,m)}_{i,k}]$ are computed, respectively.

\textbf{Lemma 4}. The average achievable rate of the typical V-UE that had access to $m$ LOS V-UEs and $n-m-1$ NLOS V-UEs  in the previous  $n-1$ steps, and is now (i.e., in $n^{th}$ step of association) connected to a BS/V-UE in the $\textit{k}^{th}$ tier is given by:


\begin{align} \label{ER}
E[R^{(n,m)}_{i,k}]&\mathop{=}\limits^{\left(a\right)}\int _{0}^{\infty }P(R^{(n,m)}_{i,k}>\rho) d\rho\\\nonumber
&=\int _{0}^{\infty }SC(2^{\frac{\rho N_{i,k}^{(n,m)}}{W} }-1)d\rho,
\end{align}
where (a) follows from $E(X)=\int _{0}^{\infty }P(X>x)dx$ for positive random variable X and $SC$ is the SINR coverage probability was given in Lemma 3.

\textbf{Lemma 5}. The average connection time of a typical V-UE which in $n^{th}$ step is associated to a BS/V-UE in the $\textit{k}^{th}$ tier, and in the previous  $n-1$ steps had access to $m$ LOS V-UEs and $n-m-1$ NLOS V-UEs is given by \eqref{eq_CT}.

\begin{figure*}[!t]
    \begin{equation}
    \label{eq_CT}
        \begin{split}
         & E[{t}^{(n,m)}_{i,k}]={B}_{i,k}t_{s}+\frac{1}{v} \cdot
            \begin{cases}
            &\int_{0}^{vt_{s} }\int _{0}^{\theta _{3} }d(x,\theta )f^{(n,m)}_{X_{k}} (x)f_{\theta } \left(\theta \right)dxd\theta+\int_{vt_{s} }^{\frac{vt_{s} }{\sin (\frac{\psi }{2} )} }\int _{\theta _{1} }^{\theta _{2} }d(x,\theta )f^{(n,m)}_{X_{k}} (x)f_{\theta } \left(\theta \right)dxd\theta. \quad{i= V2I}
            \\ \\
            &\int _{0}^{\pi }\int _{0}^{2vt_{s} \left|\cos (\beta )\right|}\int _{0}^{\theta _{3} }D(x,\theta )  f_{\theta } \left(\theta \right)f^{(n,m)}_{X_{k}}(x)f_{\beta } \left(\beta \right)d\theta dxd\beta \\&+2\int _{0}^{\pi }\int _{vt_{s} }^{\frac{2vt_{s} \left|\cos (\beta )\right|}{\sin (\frac{\psi }{2} )} }\int _{\theta _{1} }^{\theta _{2} }D(x,\theta ) f_{\theta } \left(\theta \right)f^{(n,m)}_{X_{k}} (x) f_{\beta } \left(\beta \right)d\theta dx d\beta \quad{i= V2V}
            \end{cases}
        \end{split}
    \end{equation}
\end{figure*}
\begin{proof}
Noting that the typical V-UE experiences a non-zero throughput, we know that:
\begin{align}
    \label{Et}
    E[{t}^{(n,m)}_{i,k}]&={B}_{i,k}t_{s}+(1-{B}_{i,k})E[{t_{L}}^{(n,m)}_{i,k}],
\end{align}
where ${B}_{i,k}$ was computed in Section~\ref{Sec:Beam_Sojourn} and $E[{t_{L}}^{(n,m)}_{i,k}]$ is the time that V-UE is connected before leaving the communication range of its serving node and can be computed as follows:
\begin{equation} \label{EtL}
E[{t_{L}}^{(n,m)}_{i,k}]=\frac{E(\left. d\right|d<vt_{s} )}{v} =\frac{1}{v} .\frac{E(d,d<vt_{s} )}{P(d<vt_{s} )},\\
\end{equation}
where $d(x,\theta)=\frac{x\sin (\frac{\psi }{2} )}{\sin (\theta +\frac{\psi }{2})}$ is the maximum distance that the V-UE can move before leaving the coverage range of its serving node and $v\cdot t_{s}$ is the total distance traversed by the V-UE. Note that $P(d<vt_{s} )=1-{B}_{i,k}$. Alternatively we can express the condition $d <  v t_{s} $ in terms of the distance between the V-UE and the serving node ($x$) and the motion angle of V-UE ($\theta$) as follows:

${\bullet}$  $ x <  v t_{s} $   and  $0<\theta <\theta_{3} $,

 or

${\bullet }$   $vt_{s} <x<\frac{vt_{s} }{\sin\left(\frac{\psi^{b} }{2} \right)} $, and  ${\theta_{1} <\theta <\theta _{2} } $.

%
So the connection time for V2I case can be derived as:

\begin{align} \label{EdV2I}
E(d,d<vt_{s})&=\int_{x}\int _{\theta }d(x,\theta )f^{(n,m)}_{X_{k}} (x)f_{\theta } \left(\theta \right)dxd\theta  \nonumber \\
&=\int_{0}^{vt_{s} }\int _{0}^{\theta _{3} }d(x,\theta )f^{(n,m)}_{X_{k}} (x)f_{\theta } \left(\theta \right)dxd\theta\\ \nonumber
&+\int_{vt_{s} }^{\frac{vt_{s} }{\sin (\frac{\psi }{2} )} }\int _{\theta _{1} }^{\theta _{2} }d(x,\theta )f^{(n,m)}_{X_{k}} (x)f_{\theta } \left(\theta \right)dxd\theta.
\end{align}
Similarly for the V2V case we have:
\begin{align} \label{EdV2V}
&E(d,d<2vt_{s}\left|\cos (\beta )\right|)\\ \nonumber
&=\int _{\beta }\int _{x}\int _{\theta }d(x,\theta )f_{\theta } \left(\theta \right)f^{(n,m)}_{X_{k}}(x)f_{\beta } \left(\beta \right)d\theta dx d\beta  \nonumber \\
&=2\int _{0}^{\pi }\int _{0}^{2vt_{s} \left|\cos (\beta )\right|}\int _{0}^{\theta _{3} }D(x,\theta )  f_{\theta } \left(\theta \right)f^{(n,m)}_{X_{k}}(x)f_{\beta } \left(\beta \right)d\theta dxd\beta\\ \nonumber
&+2\int _{0}^{\pi }\int _{vt_{s} }^{\frac{2vt_{s} \left|\cos (\beta )\right|}{\sin (\frac{\psi }{2} )} }\int _{\theta _{1} }^{\theta _{2} }D(x,\theta ) f_{\theta } \left(\theta \right)f^{(n,m)}_{X_{k}} (x) f_{\beta } \left(\beta \right)d\theta dx d\beta
\end{align}
where $\beta$ was defined in~\eqref{BV2Ik}. Substituting \eqref{EdV2I} and \eqref{EdV2V} in \eqref{EtL}, yields~\eqref{eq_CT}.
\end{proof}

The average throughput of the typical V-UE that had access to $m$ LOS V-UEs and $n-m-1$ NLOS V-UEs  in the previous  $n-1$ steps, and is now (i.e., in $n^{th}$ step of association) connected to a BS/V-UE in the $\textit{k}^{th}$ tier is calculated as follows:


\begin{equation} \label{Tnm}
T^{(n,m)}_{i,k} = E[R^{(n,m)}_{i,k}].E[{t}^{(n,m)}_{i,k}],
\end{equation}
where $E[R^{(n,m)}_{i,k}]$ and $E[{t}^{(n,m)}_{i,k}]$ were derived in the above two Lemmas. So the average achievable throughput for the $\textit{i}^{th}$ case is given by:
\begin{equation} \label{Ti}
T_{i}=\sum _{k}\sum _{n=1}^{\infty}\sum _{m=0}^{n-1}P_{i,k}^{(n,m)}T_{i,k}^{(n,m)} \left(\tau \right).
\end{equation}
and finally  the total average throughput of V-UE during one time slot is as follows:
\begin{equation} \label{T}
T=\sum _{i}P_{i}T_{i}\left(\tau \right),
\end{equation}
where $P_{i}$ is the occurrence probability of each three cases defined in \eqref{Pcase}.

The average delay experienced by the typical V-UE (in terms of the number of time slots)  is calculated through dividing  the content size of the requested file to the average throughput of the connection as:
\begin{equation} \label{D}
D(t_{s})=(1-p_{h})\frac{S}{T}.
\end{equation}
Note that for the Local case (with occurance probability $p_{h}$) the delay is zero.


\begin{table}[b]
\scriptsize
\caption{Channel parameters as specified in~\cite{TR38.901}}
\centering
\begin{tabular}{| p{0.5in} | p{1.2in} | p{1in} |}
\hline
{\bf Notation} & {\bf Parameter}					& {\bf Value} 					\\ \hline\hline
$f_c$, $W$ 		& Carrier frequency, Bandwidth			& $ 28$ GHz, $ 400 $ MHz 				\\ \hline
$\sigma^2$ 		&Thermal noise				& $ -174 $ dBm/Hz			\\ \hline
$\nu$			& Vehicular speed			& [0,120] km/h					\\ \hline
$K_{n} , N_c$			& Cache size, total number of contents			& 	[0,20], 100			\\ \hline
$ \lambda^{b}, \lambda ^u$			& BSs and V-UE density			& 10/${km}^{2}$, 200/${km}^{2}$				\\ \hline
${P}^{b}, {P}^{u}$		& TX power of BS and V-UE  			& 30 dbm, 23 dBm			\\ \hline
$ g_{M}^b, g_{m}^b, {\psi}^{b}$			& Main lobe gain, side lobe gain and 3 dB beamwidth for BS 			& 	18 dBi, -2 dBi, 10$^\circ$			\\ \hline
$ g_{M}^u, g_{m}^u, {\psi}^{u} $			& Main lobe gain, side lobe gain and 3 dB beamwidth for V- UE 			& 	12 dBi, -10 dBi, 30$^\circ$			\\ \hline
$\alpha _{L^u}, \alpha_{N^u},$ $\alpha_{L^b}, \alpha_{N^b} $			& Pathloss exponent for LOS BSs, NLOS BSs, LOS V-UE and NLOS V-UE 			& 2, 4, 2, 4				\\ \hline
$ a_{los}^u$, $a_{los}^b$			& Inverse of average LOS radius for V-UEs and BSs			& $0.033 m^{-1}, 0.0149m^{-1}$				\\ \hline
$ \tau$, $\rho $			& SINR and Rate threshold			& [-20,40]dB, [0.01,5]Gbps				\\ \hline
$\zeta_{k} $			& attenuation constant			& 0.45				\\ \hline
$ t_s$			& Time slot			& 1 s				\\ \hline
\end{tabular}
\label{tb:sim_para}
\end{table}

\section{Evaluation}
\label{s:eval}

In this section, we analyze caching in V2X mmWave systems by studying the impact of base station/V-UE density, beamwidth, speed, and cache size on the network performance. Note that the analytical results are numerically calculated based on the expressions derived in Section~\ref{s:algo}, whereas the Monte Carlo simulations are designed based on the 3GPP evaluation guidelines in~\cite{TR38.901}. To demonstrate the accuracy of the analytical expressions, we also report the results from the Monte Carlo simulations.
Table~\ref{tb:sim_para} summarizes the notations used in this paper together with the default values employed in the simulations and numerical evaluations.

\begin{figure}[t!]
\centering
	\begin{subfigure}{0.245\textwidth}
		\includegraphics[width=\columnwidth]{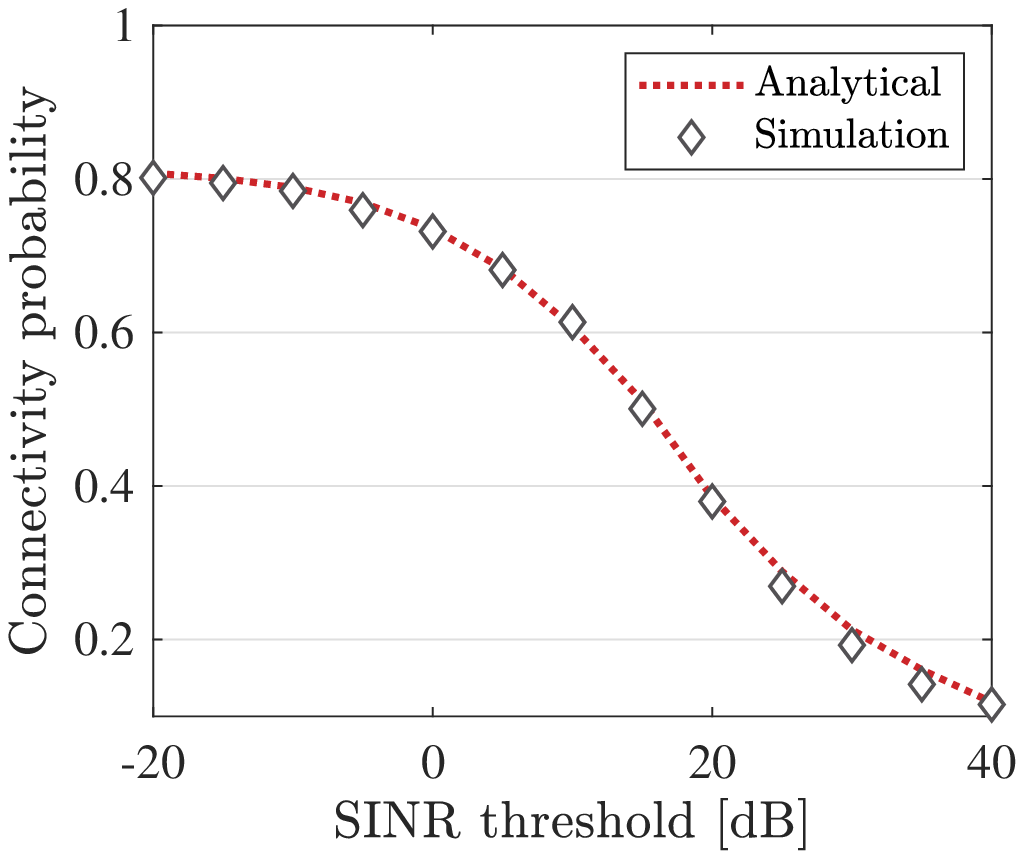}
		\caption{}		
		\label{fig:CoverageP-vs-SINR}
	\end{subfigure}
	\hspace{-3mm}
	\begin{subfigure}{0.245\textwidth}
		\includegraphics[width=\columnwidth]{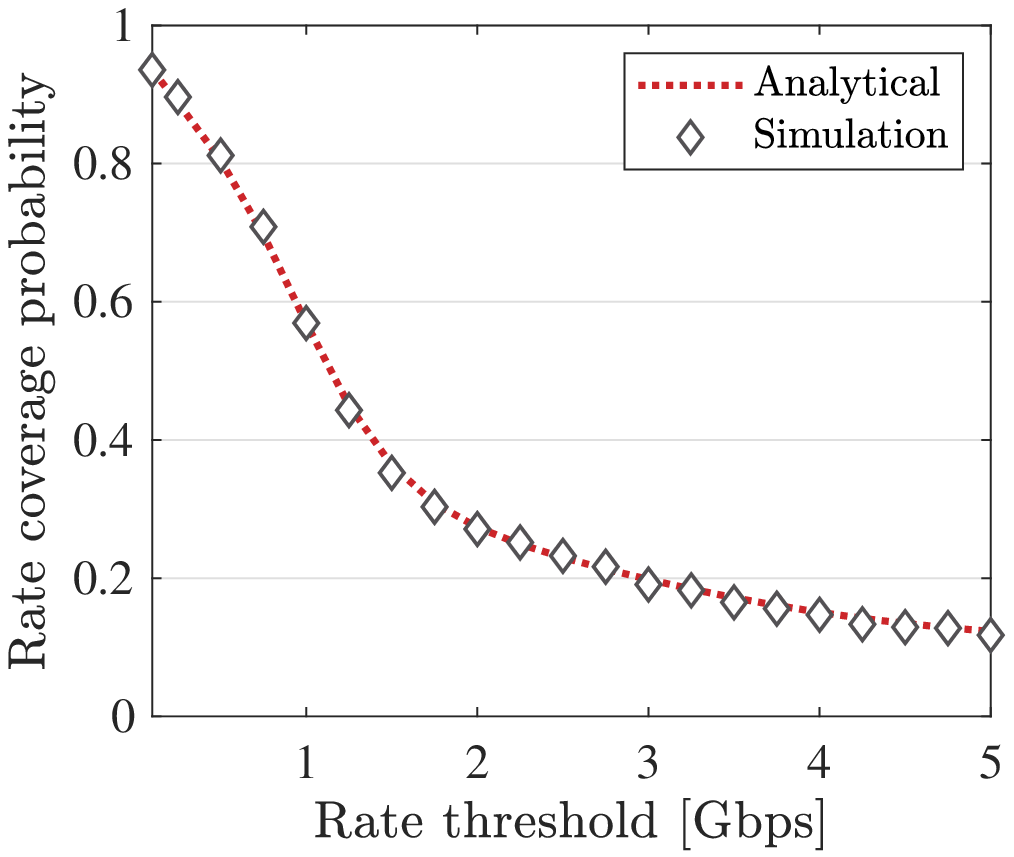}
		\caption{}		
		\label{fig:rateCoverageP-vs-Rate}
	\end{subfigure}
		\caption{The impact of SINR and rate thresholds.}
		\label{fig:thresholds}
\end{figure}

\subsection{Overall system performance}

In Fig.~\ref{fig:thresholds}, we briefly report the connectivity probability versus the SINR threshold and the rate coverage probability versus the rate threshold. These results, which are commonly presented in stochastic geometry papers, aim at verifying the derivations since the expected behavior is a priori known. 
In Fig.~\ref{fig:CoverageP-vs-SINR}, we can see that the connectivity probability reduces monotonically as the SINR threshold increases.
 This behavior is expected since the number of V-UEs under coverage effectively shrinks as we increase the minimum SINR threshold for successful decoding. For the same reason, we observe a monotonic decrease in the rate coverage probability in Fig.~\ref{fig:rateCoverageP-vs-Rate} since achieving higher rate will essentially translate into setting higher SINR thresholds.

\subsection{Impact of speed and beamwidth}
\label{ss:speed}

As shown in \eqref{PCT}, the connectivity probability depends on the SINR coverage probability and beam sojourn probability which themselves are affected by the coverage of the typical V-UE by a BS/V-UE, the quality of that connection, probability of maintaining the beam alignment within one time slot given the vehicles speed and its moving trajectory.  In Fig.~\ref{fig:speed}, we observe that the connectivity probability decreases as the speed $ v $ increases. Note that increasing the V-UE speed leads to the reduction of the beam sojourn probability because the V-UE effectively traverses the beam coverage area faster.  

While the above is expected, {\it predicting the impact of beamwidth at the BS and V-UE is not trivial} because increasing beamwidth: $ ( i ) $  increases the beam sojourn probability (the wider the beam, the larger the coverage area); and $ ( ii ) $ intensifies the interference which could potentially reduce the SINR coverage probability. Furthermore, it is unclear whether increasing the beamwidth at the BSs is more effective or that of the V-UEs.

Fig.~\ref{fig:speed} sheds light on these ambiguities. Firstly, we observe that wider beamwidths improve the connectivity probability, hence, confirming that the positive impact on beam sojourn probability (i.e., beam coverage area increases) dominates the negative impact of interference. Interestingly, using a wider beamwidth at the BSs $\psi^b$ (see Fig.~\ref{fig:connect_prob_vs_speed_bsbeam}) improves the connectivity probability much more than widening the beamwidth $\psi^u$ at V-UEs (see Fig.~\ref{fig:connect_prob_vs_speed_uebeam}). This stems in the fact that the density of V-UEs is on average higher than BSs, hence the impact of interference on the connectivity probability is higher.

\begin{figure}[t!]
\centering
	\begin{subfigure}{0.245\textwidth}
		\includegraphics[width=\columnwidth]{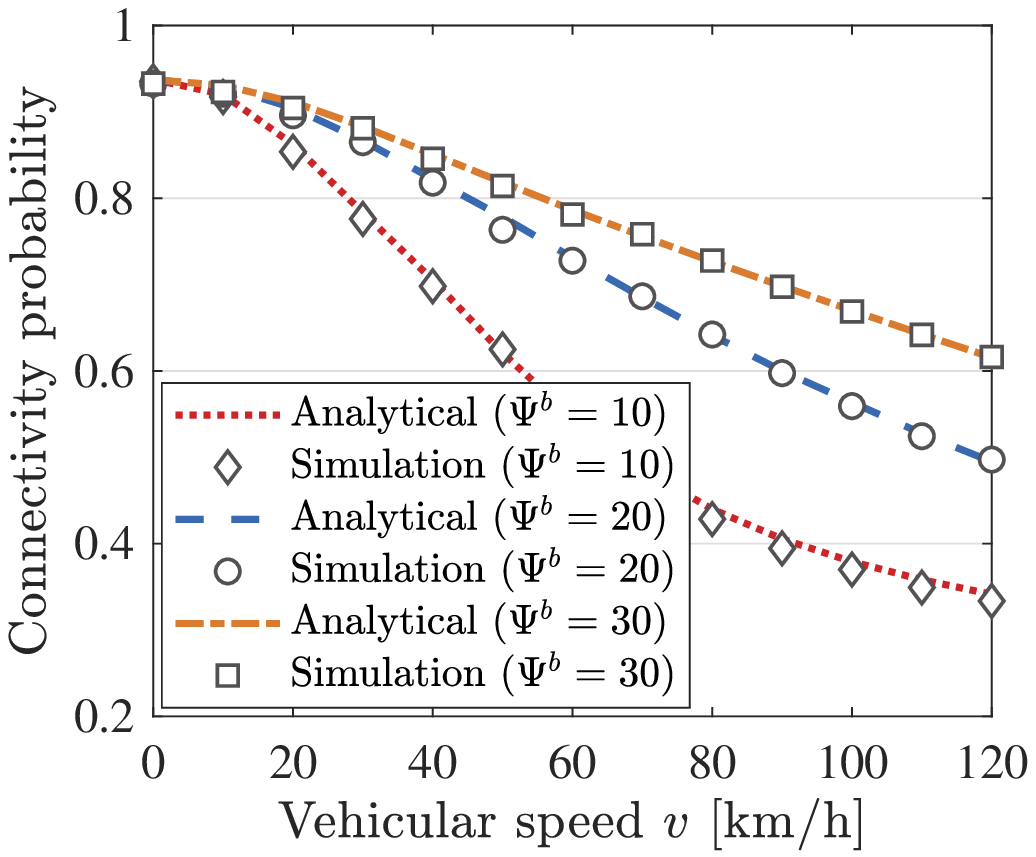}
		\caption{}
		\label{fig:connect_prob_vs_speed_bsbeam}
	\end{subfigure}
	\hspace{-3mm}	
	\begin{subfigure}{0.245\textwidth}
		\includegraphics[width=\columnwidth]{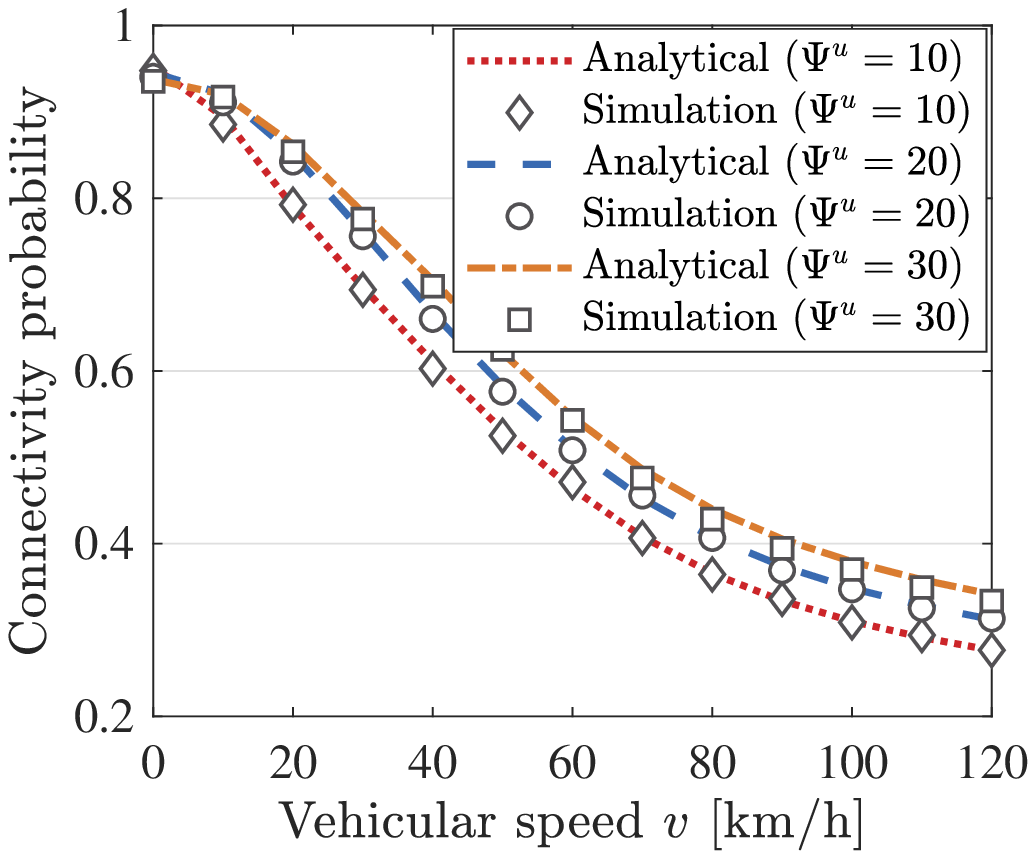}
		\caption{}
		\label{fig:connect_prob_vs_speed_uebeam}
	\end{subfigure}
		\caption{The impact of speed and beamwidth on connectivity probability.}
		\label{fig:speed}
\end{figure}

\subsection{Impact of base station and vehicle density}
\label{ss:density}
The results shown in Fig.~\ref{fig:density} demonstrates how the densities of BS $\lambda^{b}$ and V-UE $\lambda^{u}$ impact the connectivity probability.
Interestingly, the system behaves very differently in higher BS density (see Fig.~\ref{fig:connect_prob_vs_bs_density}) compared to higher V-UE density (see Fig.~\ref{fig:connect_prob_vs_ue_density}). Before describing the figures, let us discuss the dominant factors impacting them. As the network density increases: $( i ) $ the distance $x$ between the typical V-UE and its associated BS/V-UE reduces, leading to higher received power, $ ( ii ) $ the distance to other BSs/V-UEs reduces which leads to receiving higher interference, and $ ( iii ) $ the distance $x$ reduces leading to reduction of beam sojourn probability. The latter is better explained through the illustrative example in Fig.~\ref{fig:beam-sojourn}, which shows that reduction in distance $x$ leads to a smaller beam coverage area (i.e, the overlapping area between transmitter and receiver beams). Based on the dominance of the aforementioned factors, we can explain the observed trends in Fig.~\ref{fig:density}.


In Fig.~\ref{fig:connect_prob_vs_bs_density}, the connectivity probability has a maximum for a given density $\lambda^{b^*}$, after which it starts dropping. In the ascending phase, the impact of increased density (i.e., reduction of distance $x$) is higher on SINR coverage probability. In particular, up to $\lambda^{b^*}$, the improvement in the received power is much greater than that of the interference. After $\lambda^{b^*}$, however, the increments in received power is canceled by the interference caused by the transmission from nearby BSs/V-UEs. This asymptotic behavior in SINR coverage probability has been previously reported in~\cite{Wang:ij}. Note that the beam sojourn probability drops as the BS density increases (i.e., the distance $x$ decreases). However, in the ascending phase, this drop is negligible compared to the improvement of SINR coverage. In contrary, the impact beam sojourn becomes dominant in the descending phase.

In Fig.~\ref{fig:connect_prob_vs_ue_density}, we observe that the connectivity probability almost monotonically drops as the V-UE density $\lambda^{u}$ increases. Since the vehicular densities are in general much higher than the BS density, the impact of interference on SINR coverage probability becomes dominant. In addition, the beam sojourn probability also drops as distance $x$ shrinks. The combination of these two factors results in the observed trend in the figure.

Note narrower beamwidths ($\psi^{b}$ and $\psi^{u}$) result in lower connectivity probability. This also stems in the drop in the beam sojourn probability. As the beamwidth reduces, the overlapping beam coverage area decreases (see Fig.~\ref{fig:beam-sojourn}). Effectively, reducing beamwidth has the same impact as decreasing the distance $x$.

\begin{figure}[t!]
\centering
	\begin{subfigure}{0.245\textwidth}
		\includegraphics[width=\columnwidth]{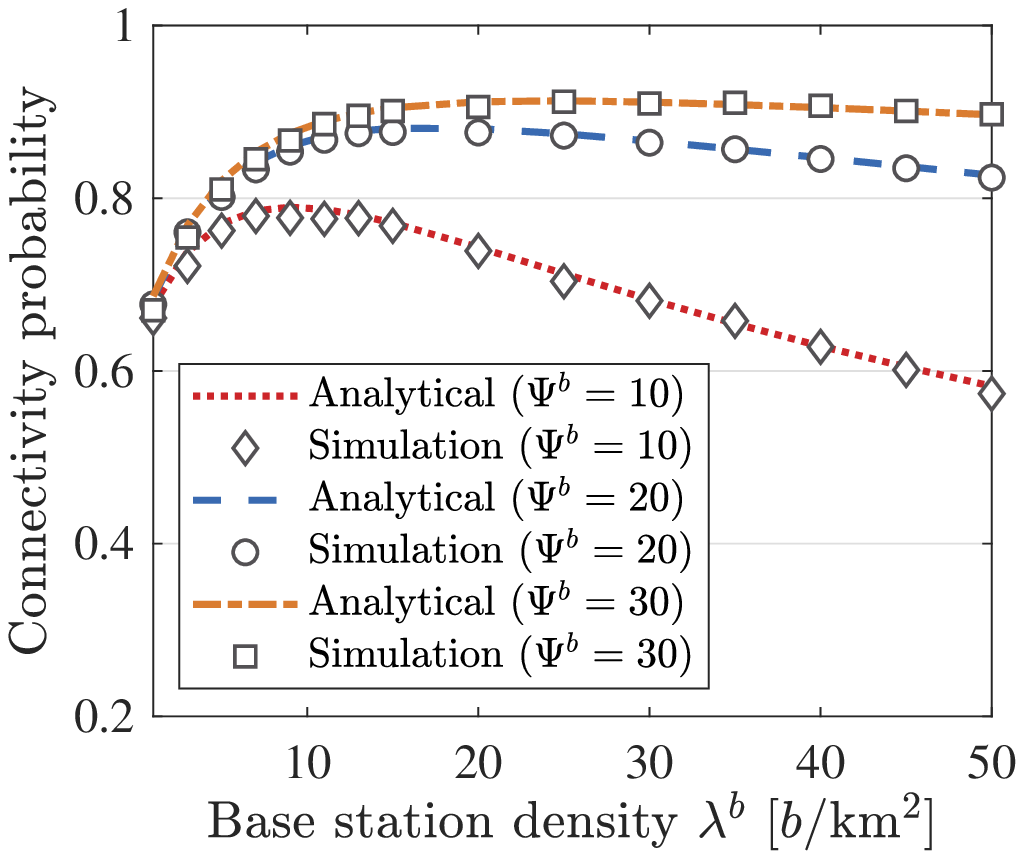}
		\caption{}
		\label{fig:connect_prob_vs_bs_density}
	\end{subfigure}
	\hspace{-3mm}
	\begin{subfigure}{0.245\textwidth}
		\includegraphics[width=\columnwidth]{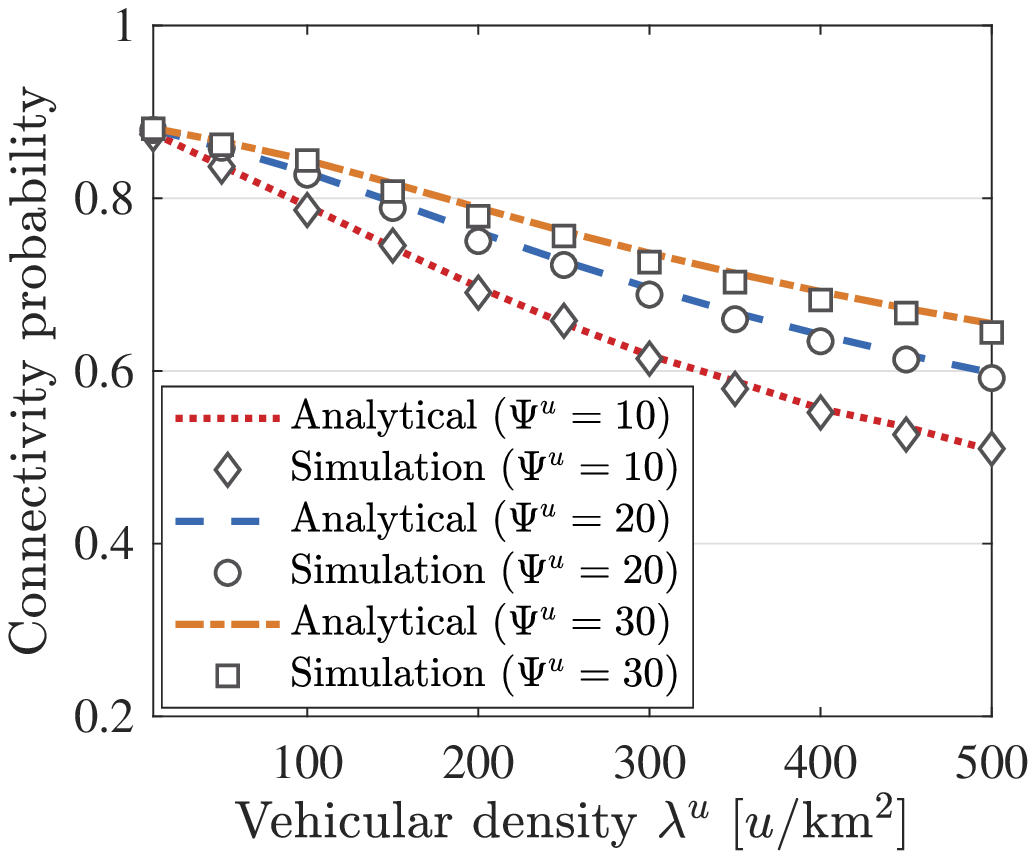}
		\caption{}
		\label{fig:connect_prob_vs_ue_density}
	\end{subfigure}
		\caption{The impact of density and beamwidth on the connectivity probability.}
		\label{fig:density}
\end{figure}

\begin{figure}[t!]
\centering
	\begin{subfigure}{0.245\textwidth}
		\includegraphics[width=\columnwidth]{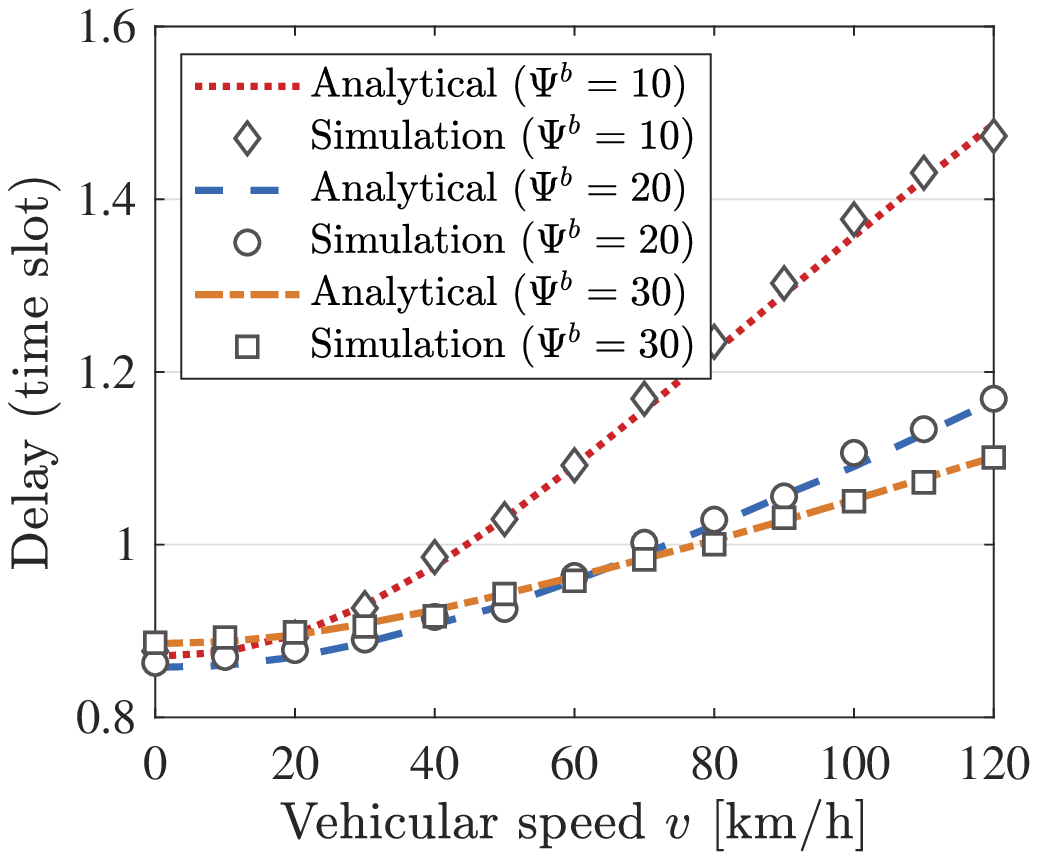}
		\caption{}
		\label{fig:delay_speed_bs_beam}
	\end{subfigure}
	\hspace{-3mm}
	\begin{subfigure}{0.245\textwidth}
		\includegraphics[width=\columnwidth]{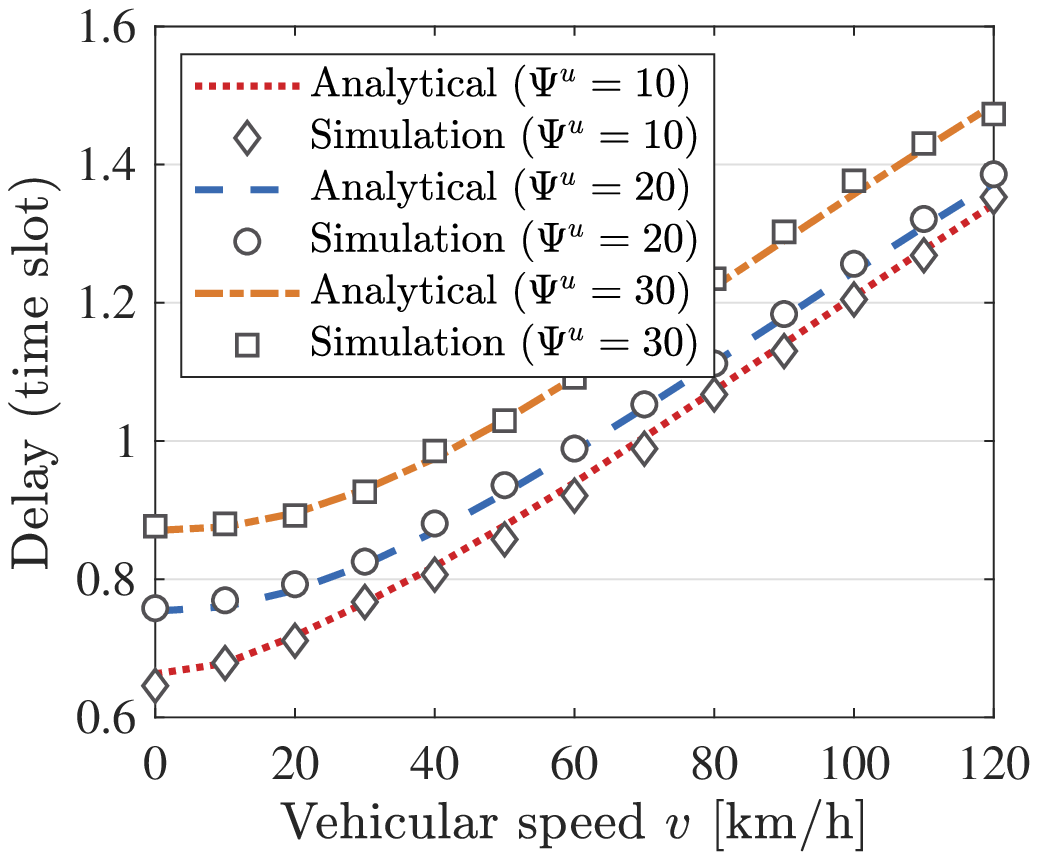}
		\caption{}
		\label{fig:delay_speed_ue_beam}
	\end{subfigure}
		\caption{The impact of speed and beamwidth on the average delay.}
		\label{fig:delay_speed}
\end{figure}

As we know the delay in receiving the desired content inversely depends on both the rate coverage probability and connection time.
In Fig.~\ref{fig:delay_speed},  the average delay is plotted versus the speed of V-UEs. We can see that the delay increases monotonically with the speed. This behavior is expected since V-UEs with higher speed experiences lower beam sojourn time  and as a result lower connectivity probability, which in turn reduces the connection time. Also as previously shown in Fig.~\ref{fig:speed}, exploiting a wider beamwidth at the BSs  ($\psi^b$) improves the connectivity probability much more than widening the beamwidth  of V-UEs ($\psi^u$). We observe similar behavior for the average delay.

\begin{figure}[t!]
\centering
	\begin{subfigure}{0.245\textwidth}
		\includegraphics[width=\columnwidth]{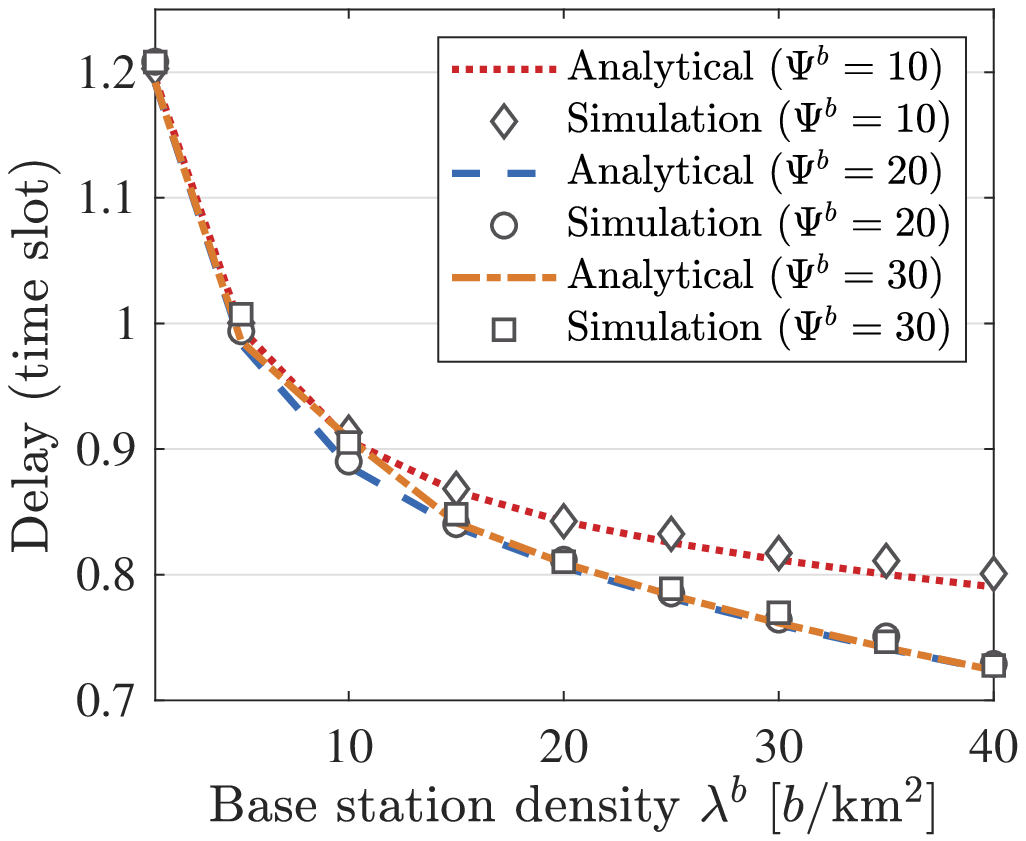}
		\caption{}
		\label{fig:delay_density_bs_beam}
	\end{subfigure}
	\hspace{-3mm}
	\begin{subfigure}{0.245\textwidth}
		\includegraphics[width=\columnwidth]{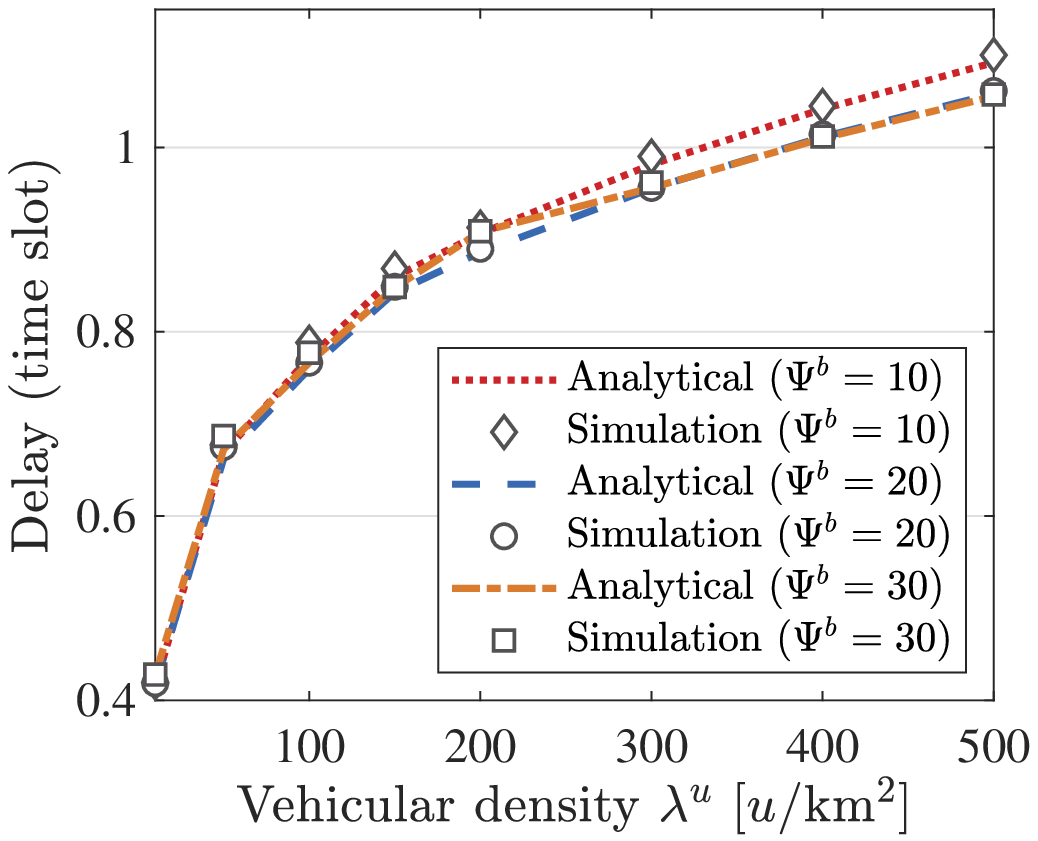}
		\caption{}
		\label{fig:delay_density_ue_beam}
	\end{subfigure}
		\caption{The impact of density and beamwidth on the average delay.}
		\label{fig:delay_density}
\end{figure}

In Fig.~\ref{fig:delay_density}, we investigate the impact of BS and V-UE densities on the delay. Comparing  Fig.~\ref{fig:delay_density_bs_beam} and Fig.~\ref{fig:delay_density_ue_beam}, we observe that densification of BSs and V-UEs impacts the delay performance differently. This behavior was also observed  in Fig.~\ref{fig:density}. In essence, increasing the the density of V-UEs has a more prominent impact on interference which leads to lower data rate and eventually higher delay.

\begin{figure}[t!]
\centering
	\begin{subfigure}{0.245\textwidth}
		\includegraphics[width=\columnwidth]{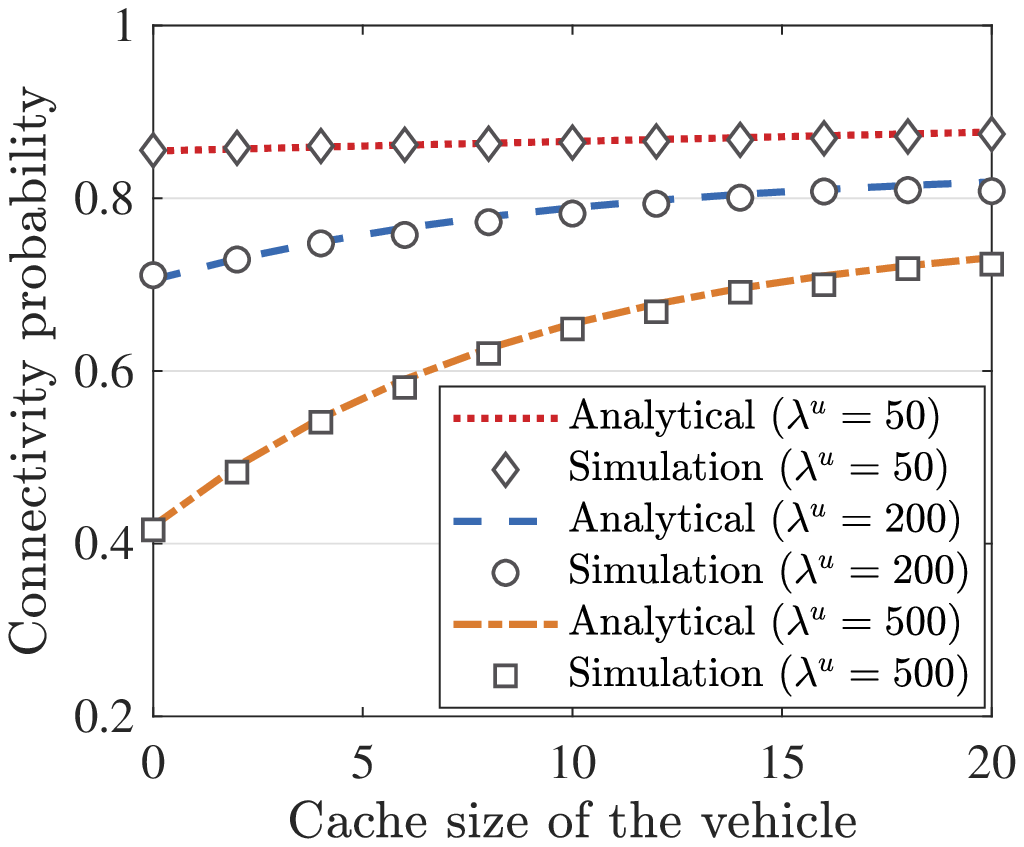}
		\caption{}		
		\label{fig:connect_prob_vs_cache_size_density}
	\end{subfigure}
	\hspace{-3mm}	
	\begin{subfigure}{0.245\textwidth}
		\includegraphics[width=\columnwidth]{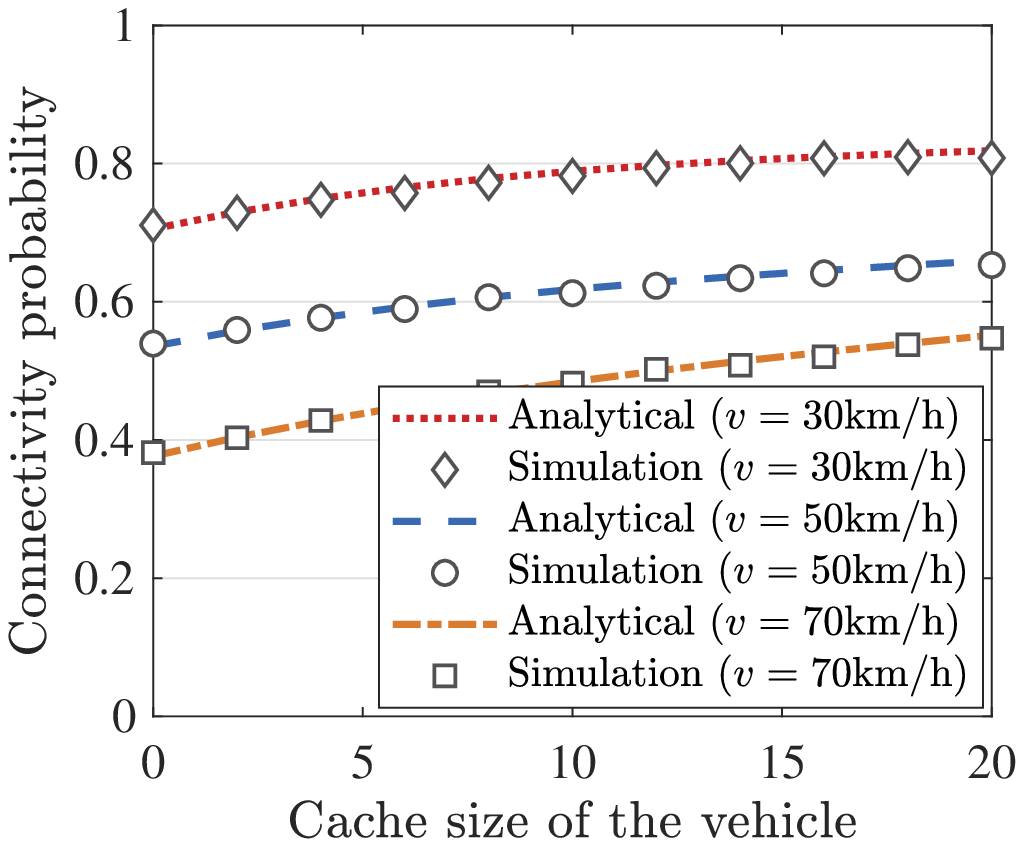}
		\caption{}
		\label{fig:connect_prob_vs_cache_size_speed}
	\end{subfigure}
	\vspace{-3mm}	
		\caption{The impact of cache size on the connectivity probability.}
		\label{fig:cache}
\end{figure}

\subsection{Impact of caching strategy}
\label{ss:cache}
Fig.~\ref{fig:cache} demonstrates the effect of cache size of the V-UEs on the connectivity probability. Fig.~\ref{fig:connect_prob_vs_cache_size_density} shows increasing cache size enhances the connectivity probability. We can also observe that the impact of cache size is much higher with higher vehicular densities. Since increasing the cache size enhances the probability of finding the required content from a nearby V-UE in less number of rounds (see \eqref{Pnm}).
The same trend is observed in Fig.~\ref{fig:connect_prob_vs_cache_size_speed} under varying V-UEs velocities. Note that, as discussed in Fig.~\ref{fig:speed}, the connectivity probability drops as the speed increases (due to the reduction of beam sojourn probability). The important design insights in~Fig.~\ref{fig:cache} are: $ ( i ) $ increasing the cache size at the V-UE is an effective method for combating the effect of interference in dense networks, and $ ( ii ) $ the negative impact of speed on connectivity can be partially compensated by deploying larger cache at the V-UEs.

\section{Conclusions}
\label{s:conclusions}
\vspace{-1mm}

In this paper, we provided an analytical frameworks based on stochastic geometry tools for caching in mmWave V2X systems. We consider a network in which all vehicles are equipped with antenna phased arrays and can perform beamforming for communicating in mmWave bands. We assume the popular contents (e.g., 3D maps, LIDAR information) are cached among vehicles randomly but the BS has all the contents available. The evaluation results showed that our analytical derivation is highly accurate when compared with the Monte Carlo simulations. Furthermore, the evaluations reveal interesting design insights regarding the suitable beamwidth, density of BS/V-UE, vehicular speed, and caching parameters.

As future work, we intend to use Mat\'ern hard-core point process to model the network with a non-homogeneous distribution of BSs and V-UEs. Although more challenging analytically, this is a better realization of today's networks. In addition, we plan to work towards more realistic assumptions including location-dependence of the content popularity as well as variable vehicular velocities.

\section*{Appendix}
\setcounter{equation}{0}
\renewcommand{\theequation}{\thesection.\arabic{equation}}

\subsection*{Appendix A}
The probability that the typical V-UE accesses to $\textit{n}_{k}^{th}$ closest node in tier \textit{k} in the $n^{th}$ step of association, where in the previous  $n-1$ steps were going to have access to $m$ LOS V-UEs and $n-m-1$ NLOS V-UEs, is obtained as
\begin{align} \label{GrindEQ__10_}
&{\mathbb P}\left(K^{(n)}=k,\bigcup _{j\in V_{m}}K^{(j)}=L^u,\bigcup _{j\in \bar{V}_{m}}K^{(j)}=N^u\right)\nonumber \\
&={\mathbb P}\left(\bigcup _{i\in K\backslash k}{{P}_{r}}^{(n_{i})}_{i}<{{P}_{r}}^{(n_{k})}_{k}<\bigcup _{i\in K\backslash k}{{P}_{r}}^{(n_{i}-1)}_{i}\right),\nonumber \\
& \hspace{10em} k=\{L^b,N^b,L^u,N^u\}.\nonumber
\end{align}
where $V_{m}$ is a m-members subset from the set of  $\{1,2,3,...,n-1\}$ and $Pr^{(n_{k})}_{k}$ is the user's received power from the $\textit{n}_{k}^{th}$ closest node in $\textit{k}^{th}$ tier ($n_{L^b}=n_{N^b}=1$, $n_{L^u}=m+1$ and $n_{N^u}=n-m$). Thus the association probability of tier \textit{k} is computed as:

\begin{align}
&{A_{k}}^{(n,m)}={\mathbb P}\bigg(\!\!\bigcup _{i\in K\backslash k}\!\!\!\!{P_{t}}_{i}g_{M_{i}}\ell_{i}(R^{(n_{i})}_{i})\!\!<\!{P_{t}}_{k}g_{M_{k}}\ell_{k}(R^{(n_{k})}_{k})\nonumber \\
&\hspace{14em}\!\!<\!\!\!\bigcup _{i\in K\backslash k}\!\!\!\!{P_{t}}_{i}g_{M_{i}}\ell_{i}(R^{(n_{i}-1)}_{i})\bigg) \nonumber \\
&\hspace{-.2em}=\hspace{-.2em}\!\!\int _{0}^{\infty }\!\!\!\!\! \hspace{-.2em} \prod _{i\in K\backslash k}\!\!\!\!\hspace{-.2em}{\mathbb P}\hspace{-.2em}\left(\!\!\ell_{i}(\hspace{-.1em}R^{(n_{i})}_{i}\hspace{-.1em})\!\!<\!\hspace{-.2em}\frac{{P_{t}}_{k} g_{M_{k}}}{{P_{t}}_{i}g_{M_{i}}}\ell_{k}(\hspace{-.1em}R^{(n_{k})}_{k}\hspace{-.1em})\!\!<\!\!\ell_{i}(\hspace{-.1em}R^{(n_{i}-1)}_{i}\hspace{-.1em})\!\hspace{-.2em}\right)\!\!f^{(n_{k})}_{R_{k} }\!(\hspace{-.1em}r\hspace{-.1em})\!dr \nonumber \\
&\hspace{-.2em}=\hspace{-.2em}\!\!\int _{0}^{\infty }\!\!\!\!\!\hspace{-.2em}\prod _{i\in K\backslash k}\!\!\!\!\hspace{-.2em}{\mathbb P}\hspace{-.2em}\left(\!\!R^{(n_{i})}_{i}\hspace{-.2em}\!\!<\!\!\ell^{-1}_{i}\hspace{-.3em}\left(\hspace{-.2em}\frac{{P_{t}}_{k} g_{M_{k}}}{{P_{t}}_{i} g_{M_{i}}}\ell_{k}(R^{(n_{k})}_{k})\hspace{-.3em}\right)\hspace{-.2em}\!\!<\!\!R^{(n_{i}-1)}_{i}\hspace{-.2em}\right)\!\!f^{(n_{k})}_{R_{k}}(\hspace{-.1em}r\hspace{-.1em})dr\nonumber \\
&=\!\!\int _{0}^{\infty }\hspace{-.5em}f_{R_{k}}^{\left(n_{k}\right)}\hspace{-.2em}\left(r\right)\hspace{-.7em}\prod _{i\in K\backslash k}\hspace{-.4em}\left[F_{R_{i} }^{\left(n_{i}\right)}\hspace{-.2em} \left(\Lambda _{k,i}\left(r\right)\right)-F_{R_{i}}^{\left(n_{i}-1\right)} \hspace{-.2em}\left(\Lambda _{k,i}\left(r\right)\right)\right]dr.\nonumber
\end{align}
where ${\mathbb P}(R_{i}<r)=F_{R_{i}}(r)$. $\Lambda _{k,i}(r)=\ell^{-1}_{i}\left((\frac{{P_{t}}_{k} g_{M_{k}}}{{P_{t}}_{i} g_{M_{i}}})\ell_{k}(r)\right)$ and $\ell_{k}(r)=r^{-\alpha_{k}}.e^{-\zeta_{k}r}$ is the path loss experienced through the link with distance $r$.

\subsection*{Appendix B}
 Denote $X_{k}$ as the distance between the typical V-UE to the serving node belongs to the tier $k$ in the $n^{th}$ step of association, where in the previous $n-1$ steps were going to get service from the  $m$ LOS V-UEs and $n-m-1$ NLOS V-UEs (but could not). The PDF of $X_{k}$ is given by

\begin{align}
&f_{X_{k}}^{(n,m)}\!\!\left(x\right)\!\!=\!{\mathbb P}(R_{k}^{(n_{k})}\!\!\le x|K^{(n)}\!\!=\!k,\!\!\!\bigcup_{j\in V_{m}}\!\!\!K^{(j)}\!\!=\!\!L^u\!,\!\!\!\bigcup _{j\in \bar{V}_{m}}\!\!\!K^{(j)}\!\!=\!\!N^u)\nonumber \\
&=\frac{{\mathbb P}(R_{k} \le x,K^{(n)}\!\!=\!k,\bigcup_{j\in V_{m}}\!\!\!K^{(j)}\!\!=\!\!L^u\!,\bigcup _{j\in \bar{V}_{m}}\!\!\!K^{(j)}\!\!=\!\!N^u)}{{\mathbb P}(K^{(n)}\!\!=\!k,\bigcup_{j\in V_{m}}\!\!\!K^{(j)}\!\!=\!\!L^u\!,\bigcup _{j\in \bar{V}_{m}}\!\!\!K^{(j)}\!\!=\!\!N^u})\nonumber \\
&=\!\!\frac{1}{A_{k}^{(n,m)}}{\mathbb P}\hspace{-.2em}\!\left(\hspace{-.1em}\!\!R_{k}^{(n_{k})}\!\! \hspace{-.2em}\le \hspace{-.2em} x;\!\!\!\!\bigcup _{i\in K\backslash k}\!\!\!\!R_{i}^{(n_{i})} \!\!<\Lambda _{k,i}\hspace{-.2em}\left(R_{k}^{(n_{k})}\right)\!\!<\!\!\!\!\bigcup _{i\in K\backslash k}\!\!\!\!R_{i}^{(n_{i}-1)}\!\!\right) \nonumber \\
&=\hspace{-.2em}\frac{f_{R_{k}}^{\left(n_{k}\right)} \hspace{-.2em}\left(r\right)}{A_{k}^{\left(n,m\right)} }\hspace{-.7em}\prod _{i\in K\backslash k}\hspace{-.5em}[F_{R_{i} }^{\left(n_{i}\right)} \hspace{-.2em}\left(\Lambda _{k,i}\left(r\right)\right)\hspace{-.2em}-\hspace{-.2em}F_{R_{i} }^{\left(n_{i}-1\right)}\hspace{-.2em} \left(\Lambda _{k,i}\left(r\right)\right)].\nonumber
\end{align}
where $n_{L^b}=n_{N^b}=1$, $n_{L^u}=m+1$, and $n_{N^u}=n-m$.

\subsection*{Appendix C}
In this appendix the Laplace transform of the total interference $I$ and the SINR coverage probability are calculated. The total interference is equal to
\begin{align}
I=\sum _{k\in K }I_{k}=\sum _{k}\sum _{j\in \phi _{k} \backslash j^{*}}{P_{t}}_{k} G_{k}^{\nu} h_{k,j}\ell_{k}(x)\nonumber \\
k=\{L^b,N^b,L^u,N^u\}.\nonumber
\end{align}
So the  Laplace transform for $I$ is calculated as
\begin{align}
&L_{I}\left(s\right)=E_{\phi,G,h}\left[\exp (-sI)\right]\nonumber \\
&\simeq E_{\phi,G,h}\left[\exp (-s\sum _{k}\sum _{j\in \phi _{k}}{P_{t}}_{k} G_{i}^{\nu} h_{k,j}\ell_{k}(x)\right] \nonumber \\
&=\prod _{k\in K}E_{\phi,G} \left[\prod _{j\in \phi _{k}}E_{h} [ \exp (-s{P_{t}}_{k}G_{i}^{\nu} h_{k,j}\ell_{k}(x) )]\right] \nonumber \\
&=\prod_{k\in K}\exp\left(\!\!-2\pi\lambda _{k}\sum _{i}{p_{G_{i}^{\nu}}}\!\!\!\int _{0}^{\infty }\!\!\!\frac{r{\rm P}_{k}\left(r\right)}{1+(s{P_{t}}_{k}G_{i}^{\nu}\ell_{k}(r))^{-1}}dr\right)\nonumber \\
&={\exp \left(\hspace{-.2em}-2\pi \lambda_{k}\sum _{i}p_{G_{i}^{\nu}}\sum _{k}W_{k}(\alpha_{k},\zeta_{k},s{P_{t}}_{k}G_{i}^{\nu})\right)}.\nonumber
\end{align}
Therefore, the SINR coverage probability of tier $\textit{k}$ is given by:
\begin{align}
SC_{k}\left(\tau \right)\hspace{.2em}&={\mathbb P} \left(\frac{{P_{t}}_{k} G_{0}^{\nu} h\ell_{k} \left(x\right)}{\sigma _{k}^{2} +I_{k} } >\tau \left|K=k\right. \right) \nonumber \\
&={\mathbb P}  \left(h>\frac{\tau \left(\sigma^{2} +I\right)}{T_{k} \left(x\right)} \left|K=k\right. \right) \nonumber \\
&=E_{x,I}\left[\exp (-\frac{\tau \left(\sigma^{2} +I \right)}{T_{k} \left(x \right)} )|K=k\right]\nonumber \\
&=\int _{0}^{\infty }\exp \left(-\frac{\tau \sigma _{k}^{2} }{T_{k} \left(x\right)} \right)L_{I } \left(\frac{\tau }{T_{k} \left(x\right)}\right)f_{x_{k} } \left(x\right)dx.\nonumber
\end{align}


\bibliographystyle{IEEEtran}


\small

\bibliography{biblio}

\begin{thebibliography}{10}
\providecommand{\url}[1]{#1}
\csname url@samestyle\endcsname
\providecommand{\newblock}{\relax}
\providecommand{\bibinfo}[2]{#2}
\providecommand{\BIBentrySTDinterwordspacing}{\spaceskip=0pt\relax}
\providecommand{\BIBentryALTinterwordstretchfactor}{4}
\providecommand{\BIBentryALTinterwordspacing}{\spaceskip=\fontdimen2\font plus
\BIBentryALTinterwordstretchfactor\fontdimen3\font minus
  \fontdimen4\font\relax}
\providecommand{\BIBforeignlanguage}[2]{{%
\expandafter\ifx\csname l@#1\endcsname\relax
\typeout{** WARNING: IEEEtran.bst: No hyphenation pattern has been}%
\typeout{** loaded for the language `#1'. Using the pattern for}%
\typeout{** the default language instead.}%
\else
\language=\csname l@#1\endcsname
\fi
#2}}
\providecommand{\BIBdecl}{\relax}
\BIBdecl

\bibitem{asadi2018fml}
A.~Asadi, S.~M{\"u}ller, G.~H. Sim, A.~Klein, and M.~Hollick, ``{FML: Fast
  Machine Learning for 5G mmWave Vehicular Communications},'' in \emph{IEEE
  INFOCOM}, 2018, pp. 1961--1969.

\bibitem{series2015imt}
ITU, ``{IMT Vision--Framework and Overall Objectives of the Future Development
  of IMT for 2020 and Beyond},'' \emph{ITU Recommendation, M Series}, 2015.

\bibitem{hu2018vehicle}
B.~Hu, L.~Fang, X.~Cheng, and L.~Yang, ``{In-Vehicle Caching (IV-Cache) Via
  Dynamic Distributed Storage Relay (D2SR) in Vehicular Networks},'' \emph{IEEE
  Transactions on Vehicular Technology}, vol.~68, no.~1, pp. 843--855, 2018.

\bibitem{grewe2018caching}
D.~Grewe, M.~Wagner, S.~Schildt, M.~Arumaithurai, and H.~Frey,
  ``{Caching-as-a-Service in Virtualized Caches for Information-Centric
  Connected Vehicle Environments},'' in \emph{IEEE VNC}, 2018, pp. 1--8.

\bibitem{singh2015tractable}
S.~Singh, M.~N. Kulkarni, A.~Ghosh, and J.~G. Andrews, ``{Tractable Model for
  Rate in Self-Backhauled Millimeter Wave Cellular Networks},'' \emph{IEEE
  Journal on Selected Areas in Communications}, vol.~33, no.~10, pp.
  2196--2211, 2015.

\bibitem{andrews2011tractable}
J.~G. Andrews, F.~Baccelli, and R.~K. Ganti, ``{A Tractable Approach to
  Coverage and Rate in Cellular Networks},'' \emph{IEEE Transactions on
  Communications}, vol.~59, no.~11, pp. 3122--3134, 2011.

\bibitem{bai2014coverage}
T.~Bai, A.~Alkhateeb, and R.~W. Heath, ``{Coverage and Capacity of
  Millimeter-Wave Cellular Networks},'' \emph{IEEE Communications Magazine},
  vol.~52, no.~9, pp. 70--77, 2014.

\bibitem{bai2013coverage}
T.~Bai and R.~W. Heath, ``{Coverage Analysis for Millimeter Wave Cellular
  Networks with Blockage Effects},'' in \emph{IEEE GlobalSIP}, 2013, pp.
  727--730.

\bibitem{alkhateeb2017initial}
A.~Alkhateeb, Y.-H. Nam, M.~S. Rahman, J.~Zhang, and R.~W. Heath, ``{Initial
  Beam Association in Millimeter Wave Cellular systems: Analysis and Design
  Insights},'' \emph{IEEE Transactions on Wireless Communications}, vol.~16,
  no.~5, pp. 2807--2821, 2017.

\bibitem{giordani2018coverage}
M.~Giordani, M.~Rebato, A.~Zanella, and M.~Zorzi, ``{Coverage and Connectivity
  Analysis of Millimeter Wave Vehicular Networks},'' \emph{Elsevier Ad Hoc
  Networks}, vol.~80, pp. 158--171, 2018.

\bibitem{banagar20193gpp}
\BIBentryALTinterwordspacing
M.~Banagar and H.~S. Dhillon, ``{3GPP-inspired Stochastic Geometry-based
  Mobility Model for a Drone Cellular Network},'' \emph{CoRR}, vol.
  abs/1905.00972, 2019. [Online]. Available:
  \url{http://arxiv.org/abs/1905.00972}
\BIBentrySTDinterwordspacing

\bibitem{MiticiGGB13}
M.~{Mitici}, J.~{Goseling}, M.~{de Graaf}, and R.~J. {Boucherie}, ``{Optimal
  Deployment of Caches in the Plane},'' in \emph{IEEE GlobalSIP}, 2013, pp.
  863--866.

\bibitem{KeshavarzianZT19}
I.~{Keshavarzian}, Z.~{Zeinalpour-Yazdi}, and A.~{Tadaion}, ``{Energy-Efficient
  Mobility-Aware Caching Algorithms for Clustered Small Cells in Ultra-Dense
  Networks},'' \emph{IEEE Transactions on Vehicular Technology}, vol.~68,
  no.~7, pp. 6833--6846, 2019.

\bibitem{BlaszczyszynG15}
B.~{Blaszczyszyn} and A.~{Giovanidis}, ``{Optimal Geographic Caching in
  Cellular Networks},'' in \emph{IEEE ICC}, 2015, pp. 3358--3363.

\bibitem{LiuY17}
D.~{Liu} and C.~{Yang}, ``{Caching Policy Toward Maximal Success Probability
  and Area Spectral Efficiency of Cache-Enabled HetNets},'' \emph{IEEE
  Transactions on Communications}, vol.~65, no.~6, pp. 2699--2714, 2017.

\bibitem{WangLWL19}
R.~{Wang}, R.~{Li}, P.~{Wang}, and E.~{Liu}, ``{Analysis and Optimization of
  Caching in Fog Radio Access Networks},'' \emph{IEEE Transactions on Vehicular
  Technology}, 2019.

\bibitem{yang2015analysis}
C.~Yang, Y.~Yao, Z.~Chen, and B.~Xia, ``{Analysis on Cache-enabled Wireless
  Heterogeneous Networks},'' \emph{IEEE Transactions on Wireless
  Communications}, vol.~15, no.~1, pp. 131--145, 2015.

\bibitem{GolrezaeiDM14}
N.~{Golrezaei}, A.~G. {Dimakis}, and A.~F. {Molisch}, ``{Scaling Behavior for
  Device-to-Device Communications With Distributed Caching},'' \emph{IEEE
  Transactions on Information Theory}, vol.~60, no.~7, pp. 4286--4298, July
  2014.

\bibitem{GiatsoglouNKAV17}
N.~{Giatsoglou}, K.~{Ntontin}, E.~{Kartsakli}, A.~{Antonopoulos}, and
  C.~{Verikoukis}, ``{D2D-Aware Device Caching in mmWave-Cellular Networks},''
  \emph{IEEE Journal on Selected Areas in Communications}, vol.~35, no.~9, pp.
  2025--2037, Sep. 2017.

\bibitem{AfshangDC16}
M.~{Afshang}, H.~S. {Dhillon}, and P.~H. {Joo Chong}, ``{Modeling and
  Performance Analysis of Clustered Device-to-Device Networks},'' \emph{IEEE
  Transactions on Wireless Communications}, vol.~15, no.~7, pp. 4957--4972,
  2016.

\bibitem{MaDCLMLV18}
C.~{Ma}, M.~{Ding}, H.~{Chen}, Z.~{Lin}, G.~{Mao}, Y.~{Liang}, and
  B.~{Vucetic}, ``{Socially Aware Caching Strategy in Device-to-Device
  Communication Networks},'' \emph{IEEE Transactions on Vehicular Technology},
  vol.~67, no.~5, pp. 4615--4629, 2018.

\bibitem{ChenPK17}
Z.~{Chen}, N.~{Pappas}, and M.~{Kountouris}, ``{Probabilistic Caching in
  Wireless D2D Networks: Cache Hit Optimal Versus Throughput Optimal},''
  \emph{IEEE Communications Letters}, vol.~21, no.~3, pp. 584--587, 2017.

\bibitem{ChoHC18}
Y.~J. {Cho}, K.~{Huang}, and C.~{Chae}, ``{V2X Downlink Coverage Analysis with
  a Realistic Urban Vehicular Model},'' in \emph{IEEE Globecom Workshops},
  2018, pp. 1--6.

\bibitem{tassi2017modeling}
A.~Tassi, M.~Egan, R.~J. Piechocki, and A.~Nix, ``{Modeling and Design of
  Millimeter-Wave Networks for Highway Vehicular Communication},'' \emph{IEEE
  Transactions on Vehicular Technology}, vol.~66, no.~12, pp. 10\,676--10\,691,
  2017.

\bibitem{GiordaniRZZ18}
M.~{Giordani}, M.~{Rebato}, A.~{Zanella}, and M.~{Zorzi}, ``{Coverage and
  Connectivity Analysis of Millimeter Wave Vehicular Networks},''
  \emph{Elsevier Ad Hoc Networks}, vol.~80, pp. 158 -- 171, 2018.

\bibitem{sial2018stochastic}
\BIBentryALTinterwordspacing
M.~N. Sial, Y.~Deng, J.~Ahmed, A.~Nallanathan, and M.~Dohler, ``{Stochastic
  Geometry Modeling of Cellular {V2X} Communication on Shared Uplink
  Channels},'' \emph{CoRR}, vol. abs/1804.08409, 2018. [Online]. Available:
  \url{http://arxiv.org/abs/1804.08409}
\BIBentrySTDinterwordspacing

\bibitem{Wang:ij}
Y.~Wang, K.~Venugopal, A.~F. Molisch, and R.~W. Heath, ``{MmWave
  Vehicle-to-Infrastructure Communication: Analysis of Urban Microcellular
  Networks},'' \emph{IEEE Transactions on Vehicular Technology}, vol.~67,
  no.~8, pp. 7086--7100.

\bibitem{di2015stochastic}
M.~Di~Renzo, ``{Stochastic Geometry Modeling and Analysis of Multi-tier
  Millimeter Wave Cellular Networks},'' \emph{IEEE Transactions on Wireless
  Communications}, vol.~14, no.~9, pp. 5038--5057, 2015.

\bibitem{TR38.901}
3GPP, ``{Technical Specification Group Radio Access Network: Study on Channel
  Model for Frequencies from 0.5 to 100 {GHz}},'' TR 38.901, 2017.

\end{thebibliography}


\end{document}